\documentclass[english,aps,prb,twocolumn,floatfix,showpacs]{revtex4}

\usepackage{babel}
\usepackage[utf8]{inputenc}  
\usepackage[active]{srcltx}  

\usepackage[T1]{fontenc}

\usepackage{color}
\usepackage{epsfig}
\usepackage{graphicx}
\usepackage{pstricks}

\usepackage{amsmath}
\usepackage{amssymb}
\usepackage{amsfonts}
\usepackage{amsthm}
\usepackage{bm}

\makeatletter
\@ifundefined{textcolor}{}
{%
	\definecolor{BLACK}{gray}{0}
	\definecolor{gray}{gray}{0.5}
	\definecolor{WHITE}{gray}{1}
	\definecolor{RED}{rgb}{1,0,0}
	\definecolor{GREEN}{rgb}{0,1,0}
	\definecolor{BLUE}{rgb}{0,0,1}
	\definecolor{CYAN}{cmyk}{1,0,0,0}
	\definecolor{MAGENTA}{cmyk}{0,1,0,0}
	\definecolor{YELLOW}{cmyk}{0,0,1,0}
}

\makeatother

\begin{document}
	
	\title{Exploring magneto-electric coupling through lattice distortions: insights from a pantograph model} 
	
		\author{D.\ C.\ Cabra}
	
	\affiliation{IFLySiB-CONICET and Departamento de F\'{i}sica, Universidad Nacional
		de La Plata, 
		Argentina}
	
	\author{G.\ L.\ Rossini }
	
	\affiliation{IFLySiB-CONICET and Departamento de F\'{i}sica, Universidad Nacional
		de La Plata, 
		Argentina}
	

	\begin{abstract}
	Multiferroic materials exhibit the coexistence of magnetic and electric order. 
	They are at the forefront of modern condensed matter physics due to their potential applications in next-generation technologies such as data storage, sensors, and actuators.
	Despite significant progress, understanding and optimizing the coupling mechanisms between electric polarization and magnetism remain active areas of research. 
	We review here a series of papers presenting a comprehensive numerical and theoretical exploration of a pantograph mechanism modeling magneto-electric coupling through lattice distortions in low dimensional multiferroic systems. 
	These works introduce and elaborate a microscopic model where elastic lattice distortions mediate interactions between spin 1/2 magnetic moments and electric dipoles, uncovering novel physics and functionalities. 
	The model successfully describes ubiquitous phenomena in type II improper multiferroics, particularly when dominant Ising spin components are 
	introduced through XXZ-type rotational symmetry breaking spin interactions.
	We also study more realistic extensions relevant for materials with higher spin magnetic ions and to materials where magnetic couplings draw higher dimensional lattices.
	
	\end{abstract}
	
	
	\maketitle

\tableofcontents
	
\section{Introduction\label{sec:Introduction}}

Multiferroic (MF) materials, those in
which electric and magnetic degrees of freedom are coupled, 
are a subject of current interest, not only for their
potential technological applications but also because of
the theoretical interest raised by the different unusual
properties and effects discovered over the last years,
a century later than the pioneering insight of P.\ Curie \cite{Curie-1894} and fifty years after
the first theoretical predictions and experimental realization in Cr$_2$O$_3$ \cite{Dzyaloshinskii-1960, Landau-1960, Astrov-1961}. 
They constitute a promising feature for device designs controlling magnetization with electric fields, 
or conversely electrical polarization with magnetic fields.
Current revival may be traced back to the start of this century, with the discovery of simultaneous polarization and magnetization 
in bismuth ferrite BiFeO$_3$. \cite{BiFeO}  
The model shows its success in describing the desired ubiquitous phenomena in type II improper MFs 
when dominant  Ising spin components are considered. Either by the introduction of single ion anisotropies 
or a XXZ breaking of the Rotational spin symmetry.
and  gigantic magneto-electric (ME) effects in rare earth perovskite manganites Te(Dy)MnO$_3$ \cite{Kimura-2003}.
Since then a series of exciting new materials and new microscopic descriptions have been developed (see for instance
the reviews [\onlinecite{Fiebig,CheongMostovoy,RameshSpalding-2007,Khomskii,TokuraR,Dong-2015,Khomskii2,Dagotto-2019}] and references therein).
Still, technologically useful multiferroic materials are very rare and their search constitutes an active area of research.

Among the large family of multiferroic materials known today,
there is a special class, dubbed type II MFs,
which are distinguished by the fact that the magnetic and
ferroelectric orders occur simultaneously through a cooperative transition.  
Two subclasses of these materials should still be distinguished: 
those in which a non collinear (usually spiral) magnetic order is observed
and the important subclass in which the magnetic order is collinear.

The main motivation for our approach arises from many different experiments where the
coupling between magnetic moments, elastic distortions
and electric dipoles have been observed, in particular
materials  \cite{Giovannetti, Streltsov} where multiferroicity has been linked
to magneto-elastic deformations in collinear spin models,
which in turn produce a net electric polarization.
What is most important in these materials is the extremely large ME coupling between magnetic and electrical properties,
even if the value of the electrical polarization can be rather small as compared to typical ferroelectric materials.
%
Very generally the high magneto-electric response appears to be associated to the magnetic frustration due to 
competing spin interactions leading to complex magnetic orders \cite{CheongMostovoy}.
Indeed, in most of multiferroic materials with  collinear spins
the magnetic order observed at low magnetic fields is of the ``uudd''
($\uparrow\uparrow\downarrow\downarrow$)  type along some particular line 
(see for instance  [\onlinecite{Medarde,CheongMostovoy,Khomskii}]
and references therein). 
We then focus on quasi-one dimensional materials with collinear low temperature magnetic orders.

\begin{figure}
	\begin{centering}
		\includegraphics[width=0.8 \columnwidth,keepaspectratio]{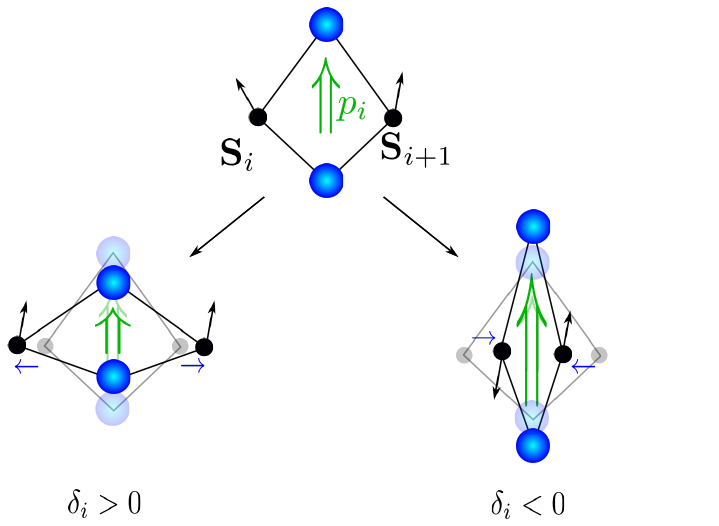}
		\par\end{centering}
	\caption{
		Schematic picture for the pantograph mechanism coupling electric dipoles
		to the lattice. Black dots represent magnetic sites and blue spheres
		represent a charge distribution giving rise to dipolar moments. Green
		double arrows represent these dipolar moments that might point up
		or down. Displacements of magnetic sites, indicated by blue arrows,
		produce lattice bond length distortions $\delta_{i}$ that modify
		the strength of local dipoles (non distorted positions are faded for reference).
		\label{fig: pantograph}
	}
\end{figure}

We  center our work on the construction and analysis of an effective microscopic model 
in which the  ME coupling is mediated by lattice distortions. 
To be precise we propose a model describing interacting spin S=1/2 magnetic ions
and interacting electric dipoles, where lattice distortions 
both affect the antiferromagnetic effective spin exchange interactions and the electric dipolar moments, 
as well as their long distance dependent interactions.
The simultaneous effects of lattice distortions on magnetic couplings and on electric dipoles reminds a pantograph mechanism, as schematically shown in Fig.\ \ref{fig: pantograph}.
They allow for a description of several transition metal materials in terms of 
almost independent chains of octahedra \cite{ANNNI-models,Dagotto3,Streltsov}. 
We should stress that the magnetic order just arises from exchange interactions, 
in contrast with non-collinear multiferroics usually modeled by spin-orbit (Dzyaloshinskii-Moriya) interactions. \cite{Khomskii2}

In several steps we first discuss a minimal model with nearest neighbors antiferromagnetic spin exchange $J_1$, 
a model where the spin-Peierls dominates the lattice dimerization distortions \cite{pantograph-1}.

\begin{figure}
	\begin{centering}
		\includegraphics[width=0.7 \columnwidth,keepaspectratio]{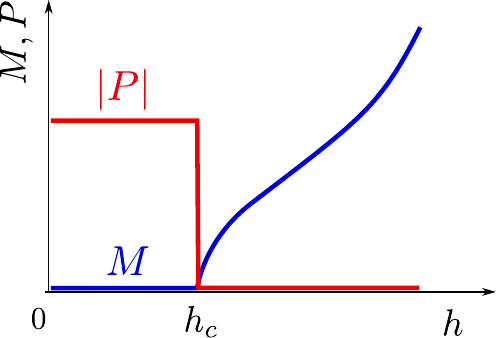}
		\par\end{centering}
	\caption{
		Schematic picture of the polarization $P$ switch-off driven by the onset of magnetization $M$ under the action of a magnetic field $h$, 
		occurring at some threshold field $h_c$. 
		The spontaneous polarization may be positive or negative along a preferred axis.
		Details in Section \ref{subsec: P-vs-M}.
		\label{fig: switch-off}
	}
\end{figure}

The very mechanism that relates magnetic order with electric polarization may be described as follows: 
the magnetic order in the absence of external fields, at low temperature, 
comes along with lattice distortions because of a gain in magnetic energy exceeding the elastic energy cost. Affecting the magnitude of the antiferroelectrically ordered dipoles,
this lattice distortion in turn produces a net ferrielectric polarization, with low enough electric energy cost or even energy gain. 
Altogether one finds a bulk polarization driven by magnetic order, that is a type II collinear multiferroic. 
The magnetic order is of course destroyed by temperature, 
but also by an external magnetic field  when the Zeeman energy gain gets larger than a finite spin gap. 
Concurrently, the lattice relaxes and the electric polarization is switched off, as illustrated in Fig.\ \ref{fig: switch-off}.
As well, an electric field high enough to produce dipole flips affects the lattice to minimize the dipole-dipole interaction energy; 
this changes the spin-spin exchange couplings and eventually alters the magnetic order.

\begin{figure}[ht!]
	\centering
	\includegraphics[width=0.7\columnwidth]{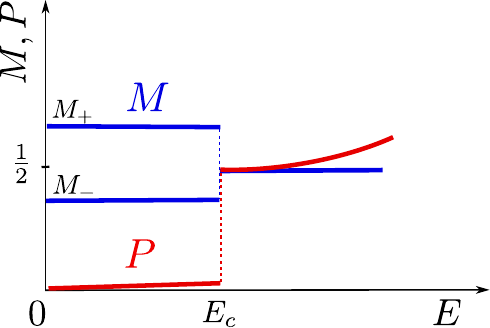} 
	\caption{Schematic picture of a magnetic response to the electric field.
		Under appropriate applied magnetic fields producing magnetizations $M^\pm$,
		an electric field  $E$ producing a first order polarization transition 	at some critical field $E_c$			
		also produces a magnetization jump from some previous incommensurate value 
		to a fractional value related to a magnetization plateau ($M=1/2$ in this picture). 
        Details in Section \ref{subsec: M-vs-P}.
	}
	\label{fig: M-jump-sketch-minimal}
\end{figure}

On the other hand, the inclusion of the electrostatic dipolar coupling introduces another playground: 
the total energy depends on the electric order, whether spontaneous or driven by external electric fields, 
and also on the lattice distortions (that locally modifies the strengths of electric dipoles and non-locally the distance between them).
In this way, as an electric field directly  drives the electric order, it also influences the elastic distortions 
and ultimately the magnetic order. We find that in some parameters range the electric field indirectly drives a jump in the magnetization, as depicted in Fig.\ \ref{fig: M-jump-sketch-minimal}

We emphasize that among other effects, this model allows for a switch-on/switch-off of the electric polarization by applying
a magnetic field, as well as  magnetization jumps induced by varying an electric field. 
These functionalities are the very key features that could lead to multiferroics based
technologies \cite{techno}.

In a second step we look for a more realistic model by the inclusion of next-to-nearest neighbors (NNN) antiferromagnetic couplings $J_2$ and 
easy-axis coupling anisotropy \cite{pantograph-2}. 
The NNN coupling introduces magnetic frustration, that is the (classical) impossibility of antiparallel spin configurations 
for any antiferromagnetically coupled pair of magnetic ions. 
It is known that for large enough $J_2/J_1$ the classical order follows the  $\uparrow\uparrow\downarrow\downarrow$  pattern (so called antiphase in the context of ANNNI models), 
where every NNN pair of magnetic ions gets antiparallel but NN pairs are parallel every two sites. 
The easy axis anisotropy reduces the transverse quantum fluctuations, making the spins $S=1/2$ behave ``more classically''. 
Altogether, these modifications allow for a magnetic ordered phase with the main features 
widely observed \cite{Medarde,CheongMostovoy,Khomskii} in collinear type II quasi one-dimensional multiferroic materials.   
Our analytical and numerical analysis of this enhanced realistic model proves that the low temperature magnetic order is still 
protected by a spin gap an that the pantograph mechanism efficiently produces 
the switch-on/switch-off of the electric polarization when a magnetic field grows above/below a finite threshold. 

\begin{figure}
	\begin{centering}
		\includegraphics[width=0.8 \columnwidth,keepaspectratio]{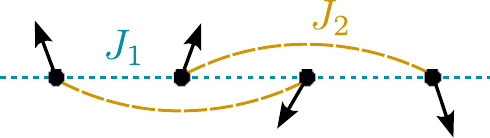}
		\par\end{centering}
	\caption{
		Schematic picture of a frustrated $J_1-J_2$ antiferromagnetic chain. 
		Large enough $J_2$ favors antiferromagnetic correlations every two sites, 
		forcing half of the first neighbors bonds to bear ferromagnetic correlations instead of
		the energetically convenient antiferromagnetic ones. Spins are drawn arbitrarily, 
		suggesting a tendency to  $\uparrow\uparrow\downarrow\downarrow$ order.
		\label{fig: J1-J2}
	}
\end{figure}
In order to understand the multiferroic transition (intertwined changes in the magnetization and electric polarization) 
we analyze the magneto-electro-elastic configurations of the system at the zero magnetization plateau, 
to be compared with the lowest magnetization excited states.
We characterize the local order, we find that the states at each side of the transition belong to different topological classes.

\begin{figure}
	\begin{centering}
		\includegraphics[width=0.8 \columnwidth,keepaspectratio]{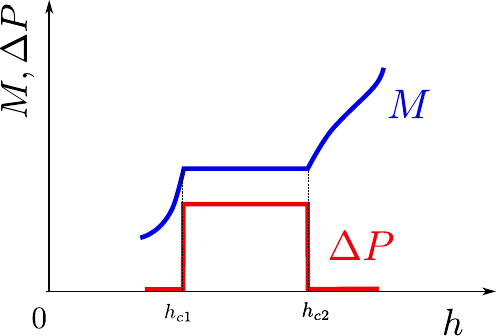}
		\par\end{centering}
	\caption{
		Generic change $\Delta P$ in the electric polarization, occurring when the magnetic field $h$
		drives the magnetic sector through a magnetization plateau.  
		A detailed example is discussed in Section \ref{sec: finite M effects}.
		\label{fig: Delta-P-Plateau-finito}
	}
\end{figure}

After understanding the nature of the magnetization onset transition, 
we further investigate the presence of other finite magnetization plateaus (spin gaps in the magnetic excitation spectrum)
where novel magnetic orders could be associated to ferroelectric properties. 
In general, a magnetic disordered phase does not generate a net polarization related to distortions; 
in contrast, entering and leaving an ordered plateau state will produce a distortion induced change in the electric polarization. 
The generic effect, 
experimentally observed for instance in R$_2$V$_2$O$_7$ (R = Ni, Co),
\cite{Ouyang-2018} is illustrated in Fig.\ \ref{fig: Delta-P-Plateau-finito}.
A particular interplay is found in the model with NNN couplings and anisotropy, where we find an interesting competition between (non-compatible) magnetic and electric orders 
at a plateau state with $M=1/3$ magnetization and $M=1/3$ polarization (with respect to saturation). 
That is, on top of  the magnetic frustration (arising from competing magnetic orders), a second frustration mechanism takes place \cite{pantograph-3}.

More realistic settings require further extensions of the one dimensional pantograph model for spin $S=1/2$.
For instance,  single ion anisotropy effects are generally present in transition metal materials, 
in relation with the distortion of the oxygen octahedra  around them. \cite{Weingart-2012} 
We recall that higher spin magnetic ions need to be considered in order to discuss the single ion anisotropy scenario
within the spin-orbit realm. \cite{Yosida-1968} 
As well, transition metal elements may form magnetic ions with magnetic moments much larger than $\mu_B/2$.
Magneto-elastic chains with $S\geq 1$ are found to present a conditional zero magnetization spin gap, 
meaning that the gap opens only when the magneto-elastic coupling exceeds a characteristic critical value.
This occurs through a different mechanism than the one observed in the $S=1/2$ spin-Peierls unconditional gap.
For integer spin values this mechanism is presumably related to the well known Haldane gap. \cite{Haldane80}
Regarding half-odd-integer spin larger than one, the existence of a spin-Peierls transition 
has been argued \cite{Affleck-Haldane-1987} but has neither been undoubtedly observed nor proven.
Instead, we have recently characterized numerically the opening of a spin gap in the  $S=3/2$ case. \cite{Spin-Peierls-2024}
From the obtained distortion patterns,  
it is clear that a pantograph coupling to dipolar degrees of freedom will generate 
a net ferrielectric polarization at zero magnetic field,
and that such polarization will be switched-off by a magnetic field above a finite threshold.

Another necessary extension is the dimensionality, 
aiming to describe multiferroic materials where the magnetic order spans a planar or bulk lattice.
Although writing down a two- or three-dimensional magneto-electro-elastic Hamiltonian looks simple,
one must notice that the highly efficient one-dimensional techniques,
either analytical like Abelian and non-Abelian bosonization
or numerical like Density Matrix Renormalization Group (DMRG),
employed in our work must be adapted to describe higher dimensional spin lattices 
One of the authors and collaborators have discussed the effects of the pantograph mechanism in the case of 
the antiferromagnetic Ising model on the  square lattice \cite{Pili-Cabra}. 
The low temperature equilibrium state is obtained by combined Montecarlo simulations of Ising variables and lattice distortions, 
taking into account the interaction energy of the associated dipolar moments.
An important result is that, in spite of the elastic cost of lattice distortions, 
the dipolar energy gain is able to stabilize the  $\uparrow \uparrow \downarrow \downarrow$ 
type E magnetic order, 
widely observed in multiferroic perovskites.

In the present review we aim to provide a comprehensive presentation of the technical details supporting the above mentioned results. 
The outline of this work is as follows: 
in Section II we present a general description of the pantograph proposal. 
In Section III  we discuss in detail the   minimal model where both  magnetic and dipolar interactions are restricted to nearest neighbors. In this simplest version of the model reproduces the switch-off of electric polarization driven by an magnetic field, concomitant with the onset of magnetization. 
In Section IV, in order to make contact with real materials, we introduce longer range interactions and crystal anisotropy effects in the model. 
We first separately summarize the changes in the electro-elastic and  magneto-elastic sectors. 
Then we present the results for the full extended model proving 
the stabilization of the ubiquitous  $\uparrow \uparrow \downarrow \downarrow$ magnetic order, 
while maintaining polarization switch-off and the other multiferroic features found in Section III. 
Section V is dedicated to analyze the interplay between electric and magnetic ordering 
in the presence of finite electric and magnetic fields, in particular when both polarization and magnetization 
are set at $1/3$ of their saturation values.
We briefly present in Section VI further extensions of this work, 
both higher spin magnetic ions and to  to higher dimensional lattices.
In Section VII we summarize and compare the findings,
highlight their significance, and discuss open questions and possible technological applications.


\section{ Microscopic model for magneto-electric coupling through lattice distortions \label{sec: Pantograph I}}

The system model under analysis describes magnetic, electric and elastic degrees of freedom in a linear chain,
in which magnetic moments and electric dipoles interact independently with the lattice that serves as the 
intermediary for an effective magneto-elastic coupling. 
Such a model requires a large number of parameters to define  the free regime of
each degree of freedom and to introduce their couplings.
After a general description we  focus on a specific
region where multiferroicity is favored and provide the particular set of parameters to be analyzed in this section. 
We aim to model a material where the magnetic order defines a preferred direction.
Notice that in such quasi-one-dimensional material the transverse interaction 
with neighboring chains significantly renormalizes the
microscopic couplings along the longitudinal direction and that,
in consequence, the parameters in our model should be
interpreted as effective ones.

\vspace{3mm}
\subsection{Magneto-elastic sector \label{sec: magneto elastic general}}

Magnetic ions positions are described as sites $i$ in a linear chain. 
Their regular positions are $x_i=i a $ where $a$ is a lattice constant but 
under distortions the ions move to $x_i+u_i$ along the chain direction, so that sites $i$ and $i+1$ 
will be separated by a distance $a+\delta_i$ with $\delta_i=u_{i+1}-u_{i}$. 
Distortions are described by the so called Holstein phonon model.
It assumes
that the most important lattice distortion contribution
is coming from optical phonons, which is a reasonable
choice given that in real materials the active magnetic
lattice is usually a sub-lattice of a more complex crystal structure. 
The relevant elastic mode is then the relative displacement 
$\delta_i$. 
Moreover, it is treated in the adiabatic approximation, 
under the assumption that phonon frequencies are much smaller than the relevant magnetic energy scale. 
The elastic energy cost of such distortions is simply given by 
\begin{equation}
	H_\text{elastic}=\frac{K}{2}\sum_{i} \delta_i ^2,
	\label{eq: H elastic}
\end{equation}
where $K$ is the lattice stiffness. 
We are interested on distortion patterns, rather than global elastic striction.
Assuming that the crystal structure regulates the average lattice spacing, 
we impose a global fixed length constraint  
\begin{equation}
	\sum_{i} \delta_i =0
	\label{eq: basic constraint}
\end{equation}

Magnetic ions themselves are represented by $S=1/2$
spin operators $\mathbf{S}_{i}$ at chain sites. 
They mainly interact through super exchange mechanisms dictated by the local crystal environment. 
A general Hamiltonian, quadratic in spin variables, can be expressed as
\begin{equation}
	H_\text{spin}=\sum_{i,j}  J_{i,j}^{\alpha,\beta} S_i^\alpha S_j^\beta -h\sum_{i}S_i^z   ,
	\label{eq: H spin generic}
\end{equation}
where upper indices indicate spin components and $h$ represents a uniform magnetic field along a preferred direction $z$.
The key point is that super exchange couplings $J$ are affected by ion displacements $\delta_i$.
Different situations for magnetic interactions (nearest or next-to-nearest neighbor couplings, isotropic or anisotropic interactions), 
and coupling dependence on ion displacements, are discussed in separate sections of this review.

\vspace{3mm}
\subsection{Electro-elastic sector}
The electric sector is modeled by a chain of dipolar moments $\mathbf{p}_{i}$ 
lying between magnetic ions at sites $i$ and $i+1$. 
In general they arise from parity and translational symmetry breaking 
in the local charge distribution of non-magnetic ions in the crystal unit cell.\cite{Khomskii}
In most observed type II multiferroics this is related to the magnetic ion
occupying one of two possible Jahn-Teller states determined by the crystal environment.
The environment is naturally affected by elastic distortions $\delta_i$, 
which may determine changes in charge distribution as well as 
the energy level or hybridization of electron orbitals bridging the super-exchange magnetic couplings. 
In brief, as the magnetic ions change their positions 
the presence, strength and orientation of dipolar moments may also change. 

It could happen that no local dipolar moment is present in the absence of distortions, 
in this case we would describe the arising dipoles by a magnitude proportional to $\delta_i$ 
and orientation along an appropriate axis. 
For some other materials a local dipolar moment might exist prior to distortions, 
along a given axis $\hat{\mathbf{e}}$
(see for instance [\onlinecite{Tagantsev-2013}]).
While this second case is our main interest, the first one is also considered in Section \ref{sec: Pili}.

The ferroelectric effects can be measured in several ways, 
most easily through changes in the electric permittivity but also in the electrical susceptibility or the 
local or net polarization. Our approach makes direct contact with polarization properties.
Related, it is worth to recall that the measurable quantity in crystals is not the absolute polarization 
but the polarization change between different states of the same compound.\cite{Resta-1993}


For definiteness we will assume that the undistorted lattice hosts electric dipoles amid magnetic ions, 
with a natural magnitude $p_0$ and a preferred axis $\hat{\mathbf{e}}$ oriented perpendicular to the chain
(the polar direction is unimportant when the magnetic sector is rotational invariant).  
Under distortions $\delta_i$ the local dipole magnitudes are naturally modified.
This is modeled in a linear approximation by $\mathbf{p}_i=p_{i}(\sigma_{i},\delta_{i})\hat{\mathbf{e}}$ with a component
\begin{equation}
	p_{i}(\sigma_{i},\delta_{i})=p_{0}\left(1-\beta\delta_{i}\right)2\sigma_{i}.
	\label{eq: beta}
\end{equation}
Here $\sigma_{i}=\pm 1/2$ is an Ising variable for the orientation of the dipole along its axis, $p_{0}$ is the dipolar
moment magnitude in the absence of distortions, and $\beta$ will be called
the (dimensionful) dipole-elastic coupling. Notice that $\beta>0$  makes dipolar moments larger as neighboring magnetic sites become closer. 
This we call the  pantograph mechanism (the name has been used  before in [\onlinecite{Jaime-2006,Yao-2008}]) as depicted in Fig.\ \ref{fig: pantograph}.  
The mechanism encodes the interaction between electric dipoles and elastic degrees of freedom. 
Comet rhomboids in the picture represent, without loss of generality, the actual parity breaking crystal environment of magnetic ions.

For a given distribution of distortions $\delta_i$ and dipoles $p_{i}(\sigma_{i},\delta_{i})$ the system acquires a bulk polarization
\begin{equation}
	P \equiv
	\frac{1}{N_s}\sum_{i=1}^{Ns} p_{i}(\sigma_{i},\delta_{i})
	= \frac{1}{N_s}\sum_{i=1}^{Ns} p_0 (1-\beta \delta_i) 2\sigma_i,
	\label{eq: polarization definition}
\end{equation}
where $N_s$ is the chain length (number of sites).

Electric dipolar momenta are considered to interact with each other, at a relevant energy scale, in a phenomenological way. 
Such interaction is  eventually determined by long range dipole-dipole interactions and/or elastic relations 
between deformations of charged and intermediate ions in the crystal \cite{Spaldin-book}. 
For the sake of definiteness we consider a Coulomb long range dipole-dipole interaction coupling 
decaying with the cube of the dipole separation,
\begin{equation}
	\lambda_D \frac{\mathbf{p}_i \cdot \mathbf{p}_j-3(\mathbf{p}_i \cdot \hat{x})(\mathbf{p}_j \cdot \hat{x})}{|x_j-x_i|^3}
	\label{eq: dipolar interaction full}
\end{equation}
which in the present geometry only contributes with the product of the transverse components $p_i$. 
Regarding the distance decay,  notice that dipoles $p_{i}$ and $p_{i+1}$ are separated by a distance $a+\eta_i$, 
where $\eta_i=(\delta_{i}+\delta_{i+1})/2$ is the distortion of the distance between adjacent dipoles. 
The electric energy of a given configuration of dipoles coupled to distortions is given by 
\begin{eqnarray}
	H_{\text{dipole}}^\text{(full range)}&=&\lambda_{D}\sum_{i}\left(
	\frac{p_{i}(\sigma_{i},\delta_{i})p_{i+1}(\sigma_{i+1},\delta_{i+1})}{\left(a+\eta_{i} \right)^{3}} \right. \nonumber \\ 
	&+& \left. \frac{p_{i}(\sigma_{i},\delta_{i})p_{i+2}(\sigma_{i+2},\delta_{i+2})}{\left(2a+ \eta_{i} + \eta_{i+1}\right)^{3}} +\cdots\right)  \nonumber \\
	&-& E \sum p_{i}(\sigma_{i},\delta_{i}) 
	\label{eq: H dip full}
\end{eqnarray}
where the dots represent longer range dipolar interactions and $E$ is an external electric field along the dipolar axis 
$\hat{\mathbf{e}}$. 
An electric field component transverse to this axis would introduce dipolar quantum fluctuations,
interesting in the context of molecular magnets \cite{Naka-2016} or the 
ferroelectric SrTiO$_3$  \cite{SrTiO} but this is out of the scope of the present review.
Though this expression may look cumbersome, we discuss it under two approximations. 
As screening effects of surrounding charges is usually important, 
we truncate the long distance interactions up to first or second neighbors.
Also, as ions displacements are very small with respect to the crystal lattice constant, 
we expand the distance decay up to linear terms in $\delta_i$.    
Thus the approximated dipole Hamiltonian $H_{\text{dipole}} $ to be used 
is quadratic in dipolar variables $\sigma_i$,
coupled by linear interaction vertices to elastic distortions. 

\subsection{Magneto-electro-elastic pantograph model}
The addition of the magnetic, electric and elastic energy gives place to the magneto-electro-elastic Hamiltonian 
\begin{equation}
	H_{MEE}= H_{\text{spin}} + H_{\text{dipole}} + H_{\text{elastic}},
	\label{eq: H_MEE}
\end{equation}
which together with the relations of spin couplings and dipolar strength with elastic distortions defines the pantograph model 
for type II multiferroic materials. In the following we analyze different scenarios for the magnetic and dipolar interactions.


\section{Minimal magneto-elastic and electro-elastic model \label{sec: minimal}}

In this Section we present the approximations that lead to a minimal pantograph model.


\subsection{Simplest approximations}

\vspace{3mm}
\paragraph*{Magnetic interactions:}

In the simplest magnetic model leading to a zero-field spin gap
the magnetic ions hold isotropic Heisenberg interactions via NN
modulated antiferromagnetic couplings $J_{1}$. 
These super-exchange couplings depend on the local crystal environment, which
in several ways may be affected by the elastic displacements of the magnetic ions. 
We assume for simplicity that the NN exchange shows a linear dependence 
on distortions 
that can be written as 
\begin{equation}
	J_1(\delta_i)=J_1 (1-\alpha \delta_i)\\
	\label{eq: J1_i}
\end{equation}
where $\alpha>0$ is called the linear (dimensionful) magneto-elastic coupling.
We disregard at this step farther neighbors magnetic interactions, 
and keep in mind the picture that positive $\alpha$ makes NN exchange stronger as magnetic ions approach each other.
Finally, we introduce the Zeeman energy associated with an external magnetic field $h$. 
This field could act along an arbitrary direction, as long as the interactions are invariant global spin rotations
(global $SU(2)$ invariance). 

The magnetic sector, coupled to lattice distortions,  is then described by the Hamiltonian 
%
\begin{equation}
	H_\text{spin}^\text{(minimal)}=\sum_{i} J_1(\delta_i) \mathbf{S}_{i}\cdot \mathbf{S}_{i+1}-h\sum_i S^z_i,
	\label{eq: H spin Peierls}
\end{equation}
The model described  by $H_\text{elastic}+ H_\text{spin}^\text{(minimal)}$ is usually called a spin-Peierls system.
It  has been used previously to study quasi-one-dimensional materials in their low 
temperature ordered phase as is the case of the spin-Peierls phase in compounds like CuGeO$_3$ 
(see for instance [\onlinecite{CuGeO}]).
In the lab, the so-called quasi one-dimensional magnetic materials contain parallel magnetic chains immersed in a three-dimensional structure.
Even though the magnetic interactions are much greater in the direction of the chains than in the other ones, 
the phonons do remember the three-dimensionality of the system. 
The one dimensional Hamiltonian $H_{\text{elastic}}+ H_{\text{spin}}^\text{minimal}$ appears when a mean field approximation is used for the effective inter-chain interaction,
which in turn arises when the phonon coordinates are integrated out. 
\cite{Essler-1997, Dobry-2007}

\vspace{3mm}
\paragraph*{Electric interactions:}

On the electro-elastic sector, the simplest model is obtained from  the Hamiltonian in Eq.\ (\ref{eq: H dip full})
when the screening makes negligible dipolar interactions beyond first neighbors, so that we truncate the dipolar Hamiltonian 
to NN interactions. 
Besides, we assume that distortions are much smaller than the NN dipole separation, 
so that  the distance dependence may be expanded up to linear terms. 
In this case the dipolar energy simply reads 
\begin{eqnarray}
	H_\text{dipole}^\text{(minimal)}&=&J_e \sum_i \left[1-\left(\beta+\frac{3}{2}\right) (\delta_i+\delta_{i+1})\right]\sigma_i\sigma_{i+1}
	\nonumber\\
	&+& 2\beta\varepsilon\sum_{i}\delta_{i}\sigma_{i}  
	\label{H dip minimal}
\end{eqnarray}
where $J_e = \lambda_D (p^0)^2$ is the undistorted effective electric exchange coupling
and $\epsilon \equiv 2 p_0^2 E$ is the dimensionless electric field. 

\vspace{3mm}
\paragraph*{Minimal pantograph model:}

With the considerations above, the magneto-electro-elastic Hamiltonian in Eq.\ (\ref{eq: H_MEE}),
in the absence of external fields, 
much simplifies to
\begin{eqnarray}
H_{MEE}^\text{(minimal)}& = & J_m\sum_i (1-\alpha \delta_i)\mathbf{S}_{i} \cdot \mathbf{S}_{i+1}
+\frac{K}{2}\sum_i (\delta_{i})^2 \nonumber \\
&\!\!\!\!\!\!\!\!\!\!\!\!\!\!\!\!\! +& \!\!\!\!\!\!\!\!\!\!\!\! J_e \sum_i \left[1-\left(\beta+\frac{3}{2}\right) (\delta_i+\delta_{i+1})\right]\sigma_i\sigma_{i+1} 
\label{eq:H-MEE-minimal}
\end{eqnarray}
As we discuss below, this simple model captures some main properties of type II collinear multiferroic materials. 
We recall that the pantograph effect on dipoles encoded in Eq.\ (\ref{eq: beta}) and 
the inclusion of dipole-dipole electrostatic couplings depending on distance 
are at the root of the electro-elastic coupling mechanism.

It is interesting to notice that,
integrating out deformations, one would obtain a quartic expression
coupling directly the magnetic and electric degrees of freedom, 
similar to that proposed to describe organic molecular solids.  \cite{Naka}
In our approach we follow a different route, analyzing on the same footing the  elastic, magnetic and electric degrees of freedom.


\subsection{Analytical and numerical methods \label{Methods}}

Prior to consider the full problem, it is worth to discuss analytical results in the  electro-elastic and magneto-elastic sectors separately.

\vspace{3mm}
\subsubsection{Electro-elastic phase diagram in the presence of an electric field}

\begin{figure}[ht]
	\centering
	\includegraphics[width=0.85 \columnwidth,keepaspectratio]{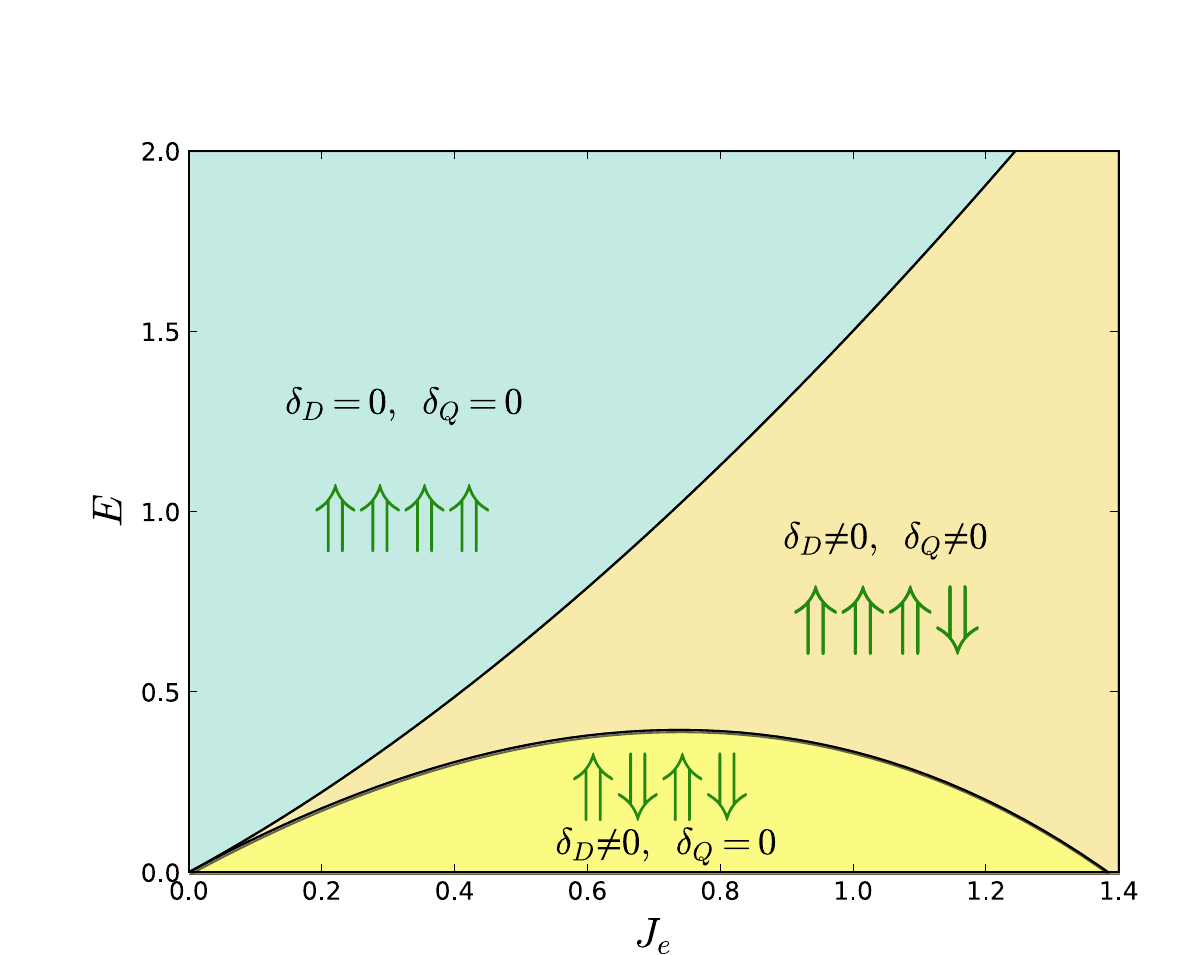}
	\caption{
	 Dipolar phases and electro-elastic distortions under an external electric field 
	 (actual data for $\alpha=0$, $\beta=0.2$ in Ref.\ [\onlinecite{pantograph-1}]). 
	 Dipolar configurations indicated by green doble arrows come along with non-trivial distortions and associated net polarization. 
	 The presence of distortion modes $\delta_D$ (dimerized) and $\delta_Q$ (quadrumerized) are indicated in each region.
	}
	\label{fig: minimal electro-elastic}
\end{figure}

The electro-elastic part of the Hamiltonian (\ref{eq:H-MEE-minimal}) (where setting $\alpha=0$ the spin sector decouples) is easily analyzed on classical grounds, for instance by Montecarlo simulations.
Distinct dipolar configurations are favored according to the electric field and the different couplings considered, 
leading to a rich phase diagram. 
We have selected the few appearing dipolar patterns (at low temperature, at most with period four) 
to analytically compute for each of them the adiabatic distortions that  minimize the 
electric plus elastic energy,  in the presence of an electric field $E$ parallel to the dipolar axis. 
Direct comparison of those energy minima gives rise to an electro-elastic phase diagram in the $E-J_e$ plane. 
We show in Fig.\ \ref{fig: minimal electro-elastic}  a typical diagram, 
for $\beta=0.2/a$; $K=1$ sets the energy scale. (revise units in fig and text).

Given the periodicity of lattice distortions, they can be analytically computed as a superposition of 
period two and/or period four harmonic distortions (formula in the figure).
The dimerized phase ({\em Dim}) with antiferroelectric $\Uparrow \Downarrow \Uparrow \Downarrow$
order appears at $E=0$ with vanishing polarization.
This phase remains until a critical field $E_{c1}$, with slightly raising alternate distortions and consequent net polarization. 
Then dipole flips occur and polarization jumps to nearly half of saturation in a quadrumerized phase ({\em Quad}) with $\Uparrow \Uparrow \Uparrow \Downarrow$ dipolar order
and period four elastic distortions (having contributions from both harmonics along this phase). With increasing field the
polarization still raises slightly, until a jump to a perfect ferroelectric order at a critical field $E_{c2}$. 
The saturated ferroelectric phase bears no distortions, recovering translational symmetry.

One should keep in mind that an electric field modifies the lattice, a fact that in turn could act on the magnetic order (when $\alpha \neq 0$). 

\vspace{3mm}
\subsubsection{Magneto-elastic spin-Peierls phenomenon} 

On the other hand,  the magneto-elastic part of the Hamiltonian (\ref{eq:H-MEE-minimal}) (setting $J_e=0$) has been extensively studied 
mainly since the discovery of CuGeO$_3$ \cite{CuGeO} and the spin-Peierls effect is well established \cite{Cross-Fisher-1979,Bray-1982}: 
in the absence of magnetic field
the system is unstable towards a lattice deformation pattern commensurate with inhomogeneous magnetic correlations 
and eventually dimerizes spontaneously. 
An efficient analysis can be made in the bosonization framework
(see [\onlinecite{Giamarchi}] for details). 
In this language the continuum expression for 
the spin energy density $\mathbf{S}_{i}\cdot\mathbf{S}_{i+1} \rightarrow \rho(x)$ reads  \cite{Haldane80}
\begin{equation}
\rho(x) =a \, \partial_x \phi
 + b :\cos(2k_F x+\sqrt{2\pi}\phi): + \cdots
\label{robos}
\end{equation}
where $\phi$ is the bosonic field, $k_F = \frac{\pi}{2}(1-M)$, $M$ is the magnetization (relative to saturation), $a$, $b$ are
$M$-dependent non-universal constants and the ellipses indicate higher harmonics. 
The magneto-elastic coupling will then provide a non vanishing relevant (cosine) operator 
when distortion modulations are commensurate with spin energy density oscillations,
opening an energy gap in the magnetic spectrum. 
This happens at zero magnetization with period two, leading to spontaneous elastic dimerization. 
The energy gain from the magnetic ground state splitting is enough to pay for the elastic energy cost of alternate distortions, 
whatever the value of the spin-phonon coupling. \cite{Cross-Fisher-1979}

\vspace{3mm}
\subsubsection{Self-consistent numerical approach \label{sec: self consistency} }

The remaining question is whether the alternate distortion accompanying magnetic interactions occurs when it implies changes in the dipolar moments.
In order to treat together all the degrees of freedom we follow a self consistent method \cite{Feiguin-etal-1997}
to find the ground state of the Hamiltonian in Eq.\ (\ref{eq:H-MEE-minimal}). 

As stated before the chains of interest are immersed in a three dimensional material 
with weak inter-chain interactions so that the effective one dimensional model 
collectively  describes a macroscopic number of chains. 
From this point of view the $\delta_i$ distortions along the chains correspond to mean field order parameters \cite{Essler-1997} 
obeying a set of self-consistent equations. 
This fact supports the validity of our approach. 

For a given configuration of dipoles $\sigma_{i}$ and a quantum state for the spins $\mathbf{S}_{i}$, 
the minimal elastic energy is obtained when distortions $\delta_{i}$ satisfy the minimal energy conditions
\begin{equation}
	\dfrac{\delta \langle H_{MEE}\rangle}{\delta_i}=0
\end{equation}
under a phenomenological fixed length condition $\sum_i \delta_{i} = 0$.

In the present case, considering the minimal pantograph Hamiltonian $H_{MEE}^\text{(minimal)}$, one explicitly gets
\begin{eqnarray}
	\label{eq: self-consistency}
	K\delta_{i}&=&
	\alpha J_{1}\langle \mathbf{S}_{i}\cdot\mathbf{S}_{i+1}\rangle
	-\beta\varepsilon\sigma_{i}\\
	&+& J_{e}\left(\beta+\frac{3}{2a}\right)\left(\sigma_{i-1}\sigma_{i}+\sigma_{i}\sigma_{i+1}\right)
	\nonumber
\end{eqnarray}
where $ \varepsilon = p_0 E$. 
We stress that, on the one hand, these self-consistent (SC) equations clearly exhibit the interplay 
between magnetic and electric degrees of freedom either collaborating or competing to produce the optimal elastic distortions. 
Each of them enters in the form of local correlations.
On the other hand the iterative procedure allows to incorporate the knowledge about the magnetic and the electric sectors separately. 

The dipolar variables $\sigma_i$ are evaluated from the decoupled electro-elastic sector. 
For zero electric field they adopt the antiferroelectric configuration $\sigma_i=\frac{(-1)^i}{2}$ (or  $\sigma_i=\frac{(-1)^{i+1}}{2}$).

The ground state for the spin system, in the dipolar background and given distortions, is obtained by the DMRG algorithm.\cite{White-DMRG}

Proving different dipolar configurations we have concluded that no dipole flips are energetically convenient. 
In practice they are kept  fixed during the iterations.

Then, we recalculate the distortions $\delta_i$ from Eq.\ (\ref{eq: self-consistency}) in order to minimize the total energy. 
This steps are iterated until energy convergence.
We have used periodic boundary conditions, and we have kept the truncation error
less than O($10^{-12}$), during up to more than 100 sweeps in the worst cases. 
This assures that errors of the DMRG computation are much smaller than symbol sizes in the shown figures.

Our computations confirm in general the robustness of the separately proposed electro-elastic and magneto-elastic mechanisms.
That is, the spin-Peierls gap remains open (active) in the presence of dipolar interactions 
and the dipolar order is stable in the presence of elastic distortions driven by the magnetic interactions.


\subsection{Polarization jump driven by a magnetic field \label{subsec: P-vs-M}}

%


The  self-consistent analysis show that present model is capable of displaying the multiferroic interplay. 
In particular, for $E=0$ and  $h=0$,
the strength of the dipoles is influenced by distortions driven by the magnetic order.
Being in the antiferroelectric Ising regime, 
the dipoles sitting in shortened bonds are enlarged in magnitude while 
those sitting in enlarged bonds are shortened in magnitude, as dictated Eq.\ (\ref{eq: beta})).
As a consequence  the magnetic frustration drives the electric subsystem to a \textit{ferrielectric} state, 
carrying a spontaneous bulk electric polarization
that can be expressed as
\begin{equation}
	P^z_{total} (h=0) \equiv \frac{1}{p^0}\sum_i p^z_i = \sum_i \sigma_i(1-\beta \delta_i)  = \pm P_{\text{sp}},
	\label{eq: net polarization minimal}
\end{equation}
where $P_{\text{sp}}=\beta \delta_D N $, 
due to the dimerized elastic distortions of amplitude $\delta_D$ associated to the spin-Peierls state,
\begin{equation}
	\delta_i = \cos(\pi  i + q \pi) \delta_D \ , \ q=0,1 \ .
		\label{eq: delta i}
\end{equation}
Such a bulk polarization, due to incomplete compensation of local dipole moments, 
has been observed in several multiferroic materials.
Some examples are AgCrS$_2$ \cite{Streltsov} 
TbMnO$_3$ \cite{TbMnO3} and TbMn$_2$O$_5$ \cite{TbMn2O5}. 
The two-fold degeneracy of the magnetic sector, and the period two dipolar configuration,  allow
to locate spin singlets (short bonds) either where dipoles point up or down. 
Then the spontaneous polarization then has two possible orientations, 
as dictated by the $\mathbb{Z}_2$ inversion symmetry of the model.

\begin{figure}[ht]
	\centering
	\includegraphics[width=0.85 \columnwidth,keepaspectratio]{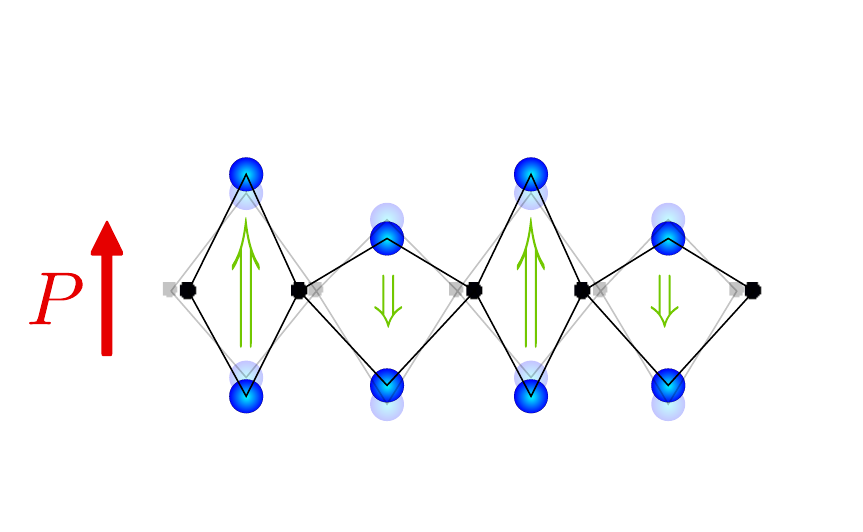} 
	\caption{
		An alternate distortion configuration  
		producing  net polarization by strength changes in an antiferroelectric 
		$\Uparrow \Downarrow \Uparrow \Downarrow$ dipolar phase
		(spins are not drawn).} 
	\label{fig: cartoon net polarization}
\end{figure}

The two possible orientations are related to the $\mathbb{Z}_2$ degeneracy,
that in turn produces a spontaneous breaking of inversion symmetry along the $z$ axis.
One of them is illustrated in the cartoon of Fig.\ \ref{fig: cartoon net polarization}.

In the presence of an electric field (not enough to produce dipole flips, see Fig. \ref{fig: minimal electro-elastic}), 
the dipole-field term in the SC equations also favors the alternation of distortions. 
But now there is an energy gain when short bonds are located where dipoles point along the field. 
In other words, an infinitesimal poling electric field breaking the $\mathbb{Z}_2$ symmetry 
is enough to select one of the otherwise degenerate electric polarization states of the system.

By increasing the magnetic field above the spin gap ($h>h_{c1}$) there occurs  
an incommensurate transition with the excitation of localized singlets  into triplets.
The  $\mathbb{Z}_2$ degeneracy of the ground state distortions has a dramatic effect on the net polarization: 
as magnetic excitations appear, distortions form regular domains interpolating between $q=0,1$, 
and the global polarization $P^z_{total}$ vanishes identically. 

Thus {\em the magnetic transition causes a complete switch-off of electrical polarization},
%
%
$P^z_{total}(h>h_{c1})  =0$. 
This simultaneous change of magnetic and dipolar orders is at the core of type II multiferroicity. 
It can be experimentally observed combining inelastic neutron scattering (spin channel) and X-ray diffraction (elastic channel) .

\begin{figure}[ht]
	\centering
	\includegraphics[width=0.75 \columnwidth,keepaspectratio]{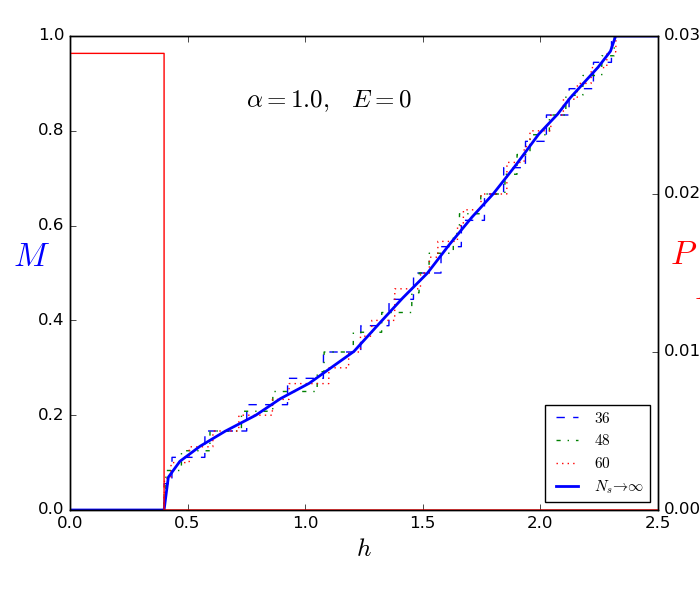} 	
	\caption{ 
		Magnetization curve (relative to saturation) and net polarization. 
A finite magnetic field is necessary to overcome the spin gap, dropping off the spontaneous polarization.
Original data in Ref.\ [\onlinecite{pantograph-1}].
} 
	\label{fig: magnetization curve E=0 minimal}
\end{figure}

The numerical results shown in Fig.\ \ref{fig: magnetization curve E=0 minimal} express the  polarization switch-off mechanism 
( $J_m=1$, $J_e=0.5$, $\alpha=1$ and $\beta=0.2$): 
they show the presence of a magnetization plateau with $M=0$ and 
a critical magnetic field $h_{c1}$ to overcome it. 
The finite net polarization at the magnetic plateau, computed from the local elastic distortions, 
drops sharply to zero as the system is magnetized.

The regularity of distortion domains, responsible for strict vanishing of the polarization, is robust because of topological reasons.
A magnetic excitation with $S_z=1$ (a magnon) on top of the $M=0$  plateau splits into a pair of solitons 
separating shifted distortion domains. 
As the solitons repel each other (an interaction with exponential decay) 
they separate as much as possible, getting equidistant in a periodic chain. 
As magnetization rises, more magnons decay into soliton pairs and repulsion makes they form a regular periodic array \cite{Lorenz_1998}.
\begin{figure}[ht]
	\centering
	\includegraphics[width=0.8 \columnwidth,keepaspectratio]{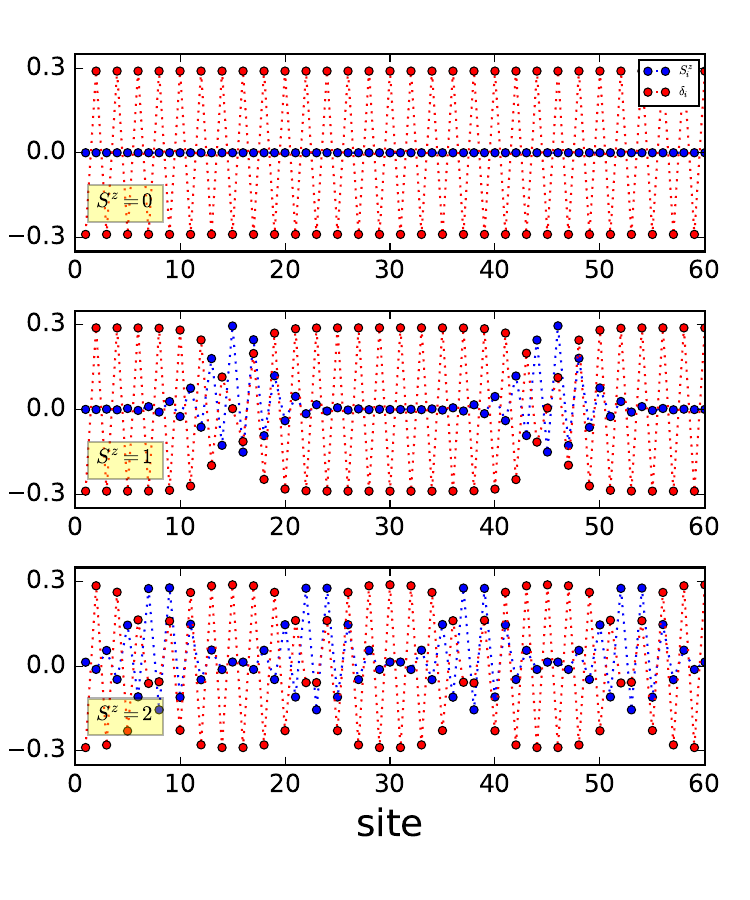} 
	\caption{Topological character of the magnetic excitations (reprinted from Ref.\ [\onlinecite{pantograph-1}]).
		Top panel: local distortions (in blue) and local spin projection (in red) computed for the $M=0$ magnetization plateau.
		Middle panel: the same quantities for the $S^z_{total}=1$ state (one spin flip). 
		Instead of a delocalized magnon, the magnetic excitation appears fractionalized and localizes 
		forming two solitonic domain walls, each one carrying $S^z=1/2$.
		Bottom panel: the same quantities for the $S^z_{total}=2$ state. Solitons proliferate as the magnetization is increased.
	} 
	\label{fig: solitons E=0 minimal}
\end{figure}

This mechanism is proven by local numerical data in Fig.\ \ref{fig: solitons E=0 minimal}, 
where the expectation value of $S_i^z$ and bond length distortions $\delta_i$ 
at the $M=0$  plateau state and lowest $M$ magnetically excited states are shown for each site and bond in a periodic lattice.
In detail, for $M=0$ distortions  alternate all along the lattice forming a single domain and  $P^z_{total}\neq 0$. 
Local spin expectation values vanish while spin-spin correlations (not shown) are dimerized, indicating the tendency to form spin singlets
located at shorter bonds.
For $S^z_{total}=1$ two equidistant domain walls appear, separating domains with twisted alternate distortion patterns
(in field theory language, topological solitons interpolating between different vacuum states). 
These domains produce opposite polarizations, so that $P^z_{total} = 0$.
The $S^z_{total}=2$ data show the same mechanism, where soliton pairs proliferate as the magnetic field is increased.


\subsection{Magnetization jump driven by an electric field \label{subsec: M-vs-P}}

\begin{figure}[ht]
	\centering
	\includegraphics[width=0.75 \columnwidth,keepaspectratio]{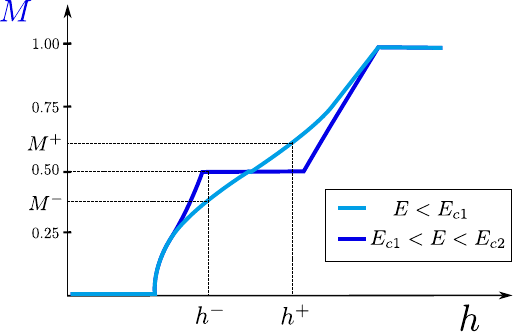} 	
	\caption{
		Magnetization curves for $E \neq 0$ (sketch from actual data in Ref.\ [\onlinecite{pantograph-1}] 
		setting $J_m=1$, $J_e=0.5$, $\alpha=1$ and $\beta=0.2$). 
		A plateau at $M=0$ is always present; for $E<E_{c1}$ this is the only plateau. 
		When $E>E_{c1}$ drives the dipolar system into a quadrumerized phase a second plateau opens at $M=1/2$.
		Magnetic fields	$h^\pm$ close to the boundaries of the $M=1/2$-plateau are marked to indicate how magnetization depends on the electric field.
}
	\label{fig: M vs H E=02 minimal electro-elastic}
\end{figure}


The presence of a small finite electric field along the dipolar direction
energetically favors the regions where dipoles point parallel to the field
(see Eq.\ (\ref{eq: H dip full})). 
Combined with a magnetic field above $h_{c1}$, the electric field enlarges such domains and shrink the others, 
providing an effective attraction that glues the soliton-antisoliton pairs. 
This damps the cancellation effect in the polarization  
and produces a net polarization along the electric field.

At some critical value $E_{c1}$ the electric field induces dipole flips, 
driving the electric subsystem to a $\Uparrow \Uparrow \Uparrow \Downarrow$ configuration
and period four elastic distortions.
We have checked numerically that this (electro-elastic) picture remains qualitatively the same 
when the distortions are coupled to the magnetic sector ($\alpha \neq 0$), 
leading to a smooth renormalization of the phase boundaries in Fig.\ \ref{fig: minimal electro-elastic}.
Now, being the distortions a superposition of 
period two  and period four  harmonics, 
the presence of a new magnetization plateau at $M=1/2$ is anticipated. 
We have computed numerically, by DMRG and self-consistency, magnetization curves in presence of electric fields.
Representative cases exhibiting plateaus are shown schematically in Fig.\ \ref{fig: M vs H E=02 minimal electro-elastic},
for  values of $E=0.2,\ 0.45$ and $\alpha= 1.0$ (actual numerical data is given in [\onlinecite{pantograph-2}]).
One observes that the plateau at $M=0$ is always present, 
while a second plateau opens at $M=1/2$ when $E>E_{c1}$ drives the dipolar system into the quadrumerized phase.
The plateau widths (a measure of the spin gap) are clearly enhanced by higher magneto-elastic coupling $\alpha$.


Because of the effects of an electric field on the magnetization just discussed above,
the minimal pantograph model presents a much desirable feature in multiferroic materials,
namely a jump in magnetization driven by electrical means.

To exhibit this, let us analyze the scenario in which both dimerized and quadrumerized phases appear as a function of $E$, 
{\it e.g.} by choosing $J_e =0.5$, $\beta = 0.2$ (see Fig. \ref{fig: minimal electro-elastic}). 
For $E_{c1} < E < E_{c2}$ the dipolar sector is quadrumerized and so is the lattice, 
which forces the magnetic sector to open a plateau at  $M=1/2$, as clearly seen 
from the numerical results in Fig.\ \ref{fig: M vs H E=02 minimal electro-elastic}. 
Choosing a background magnetic field $h^-$ at the lower boundary of this plateau, 
the magnetization will jump from some value $M^-<1/2$ to $M=1/2$ as the electric field crosses $E_{c1}$ from below; 
conversely, choosing  $h^+$ at the upper boundary 
the magnetization will jump from some value $M^+>1/2$ to $M=1/2$. This ME response is schematically depicted in Fig.\ \ref{fig: M-jump-sketch-minimal}.
Such control of magnetization by an electric field is one of the aims of multiferroic technology developments 
\cite{techno}. 
%


\subsection{Minimal pantograph model highlights}

To conclude this Section, we have proven that main multiferroic features are described by the minimal pantograph model in Eq.\ 
(\ref{eq:H-MEE-minimal}), as response to magnetic and electric fields. 
The key ingredient in the model are the elastic lattice distortions, 
separately coupled to the magnetic and to the electric degrees of freedom. 
This mediated coupling seems to be ubiquitous in magneto-electric phenomena and, 
promisingly,  
may be enhanced  by the strong influence of the lattice in multilayer multiferroics. 
Indeed,
in some cases the lattice mismatch of the layer and the substrate can generate enormous lattice
distortions and trigger giant multiferroic responses \cite{Hou-2013,Shimamoto-2017}. 

In the next Section we extend the model to look for a more realistic one, still retaining the valuable results of the minimal one. 


\section{Extended pantograph model  \label{subsec: Pantograph II}}

An important
subclass of type II collinear multiferroic materials is that presenting the
 $\uparrow \uparrow \downarrow \downarrow$ magnetic 
order at low temperatures,
that is an
arrangement of spins following a period 4 pattern $\uparrow \uparrow \downarrow \downarrow$
in one, two or the three directions of the crystal.
Such order usually appears when second neighbors antiferromagnetic interactions compete 
with the uniform or N\'eel configurations induced by nearest neighbors interactions.
This happens to be the case in quasi-one-dimensional materials like 
Ca$_{3}$CoMnO$_{6}$ \cite{Cheong-2008},
quasi-two-dimensional materials like delafossite 
AgCrS$_{2}$ \cite{Damay-2011,Streltsov} 
and also in multiferroic 
manganite perovskites with E-type antiferromagnetic order such as 
HoMnO$_{3}$ \cite{Dagotto-2006,Dong-2009}, 
ferrite perovskites such as
GdFeO$_{3}$ \cite{Tokura-2009}
and other 3D compounds such as the 
CdV$_{2}$O$_{4}$ spinel \cite{Giovannetti-2011}
or 
RNiO$_{3}$ nickelates (R=La, Pr, \dots,Lu) \cite{Catalano-2018}.
Among these $\uparrow\uparrow\downarrow\downarrow$ multiferroic materials, particular interest focuses on double perovskites such as Yb$_{2}$CoMnO$_{6}$ \cite{Blasco-2017}, 
Lu$_{2}$MnCoO$_{6}$ \cite{Batista-2011,Chikara-2016},
Er$_{2}$CoMnO$_{6}$ \cite{Oh-Oh-2019}, and R$_{2}$NiMnO$_{6}$ (R=Pr, Nd, Sm, Gd, Tb, Dy, Ho, and Er) where a giant magneto-electric effect has been reported  \cite{Zhou-2015}.
While the model aims to describe the $\uparrow\uparrow\downarrow\downarrow$ order observed along 
certain lines in those two and three dimensional multiferroic materials, 
it is interesting to notice that 
a few compounds that have been identified to become multiferroic 
do show this order in quasi-one-dimensional chains of Cu$^{2+}$ magnetic ions (S = 1/2): 
for instance
LiCuVO$_4$ \cite{6,7}, 
LiCu$_2$O$_2$ \cite{8,9,10}, 
CuCl$_2$ \cite{11}, 
CuBr$_2$ \cite{12}, 
PbCuSO$_4$(OH)$_2$ \cite{13,14}, 
CuCrO$_4$ \cite{15} 
and SrCuTe$_2$O$_6$ \cite{SrCuTeO-2016}.

In this Section we extend and generalize our previous study in several aspects. 
First and most important, we add antiferromagnetic exchange couplings between next nearest neighbors (NNN) 
reported in most of the above mentioned materials.  When the NNN coupling is strong enough we reproduce 
the experimentally observed $\uparrow\uparrow\downarrow\downarrow$ 
magnetic ordering at zero magnetic field. 
This confirms that magnetic frustration is at the root of the phenomenology observed in many materials.

Second and in order to make closer contact with experiments,  
we introduce an easy axis anisotropy that mimics the effective Ising character observed for otherwise quantum magnetic moments. 
Indeed, the  magnetic ions are immersed in crystal local fields that generally diminish their quantum character, 
making them 
behave in many materials as almost classical Ising variables. 
Good examples of this situation are the spin-ice pyrochlores \cite{Bramwell-2001}, 
with the exception being Tb based pyrochlores where Ising models seem not to suffice but quantum fluctuations have to be included 
\cite{Gingras-2014,Rau-2018,Pili-2020}.
Thus a parameter controlling the easy axis anisotropy allows for a phase diagram covering the ``quantum'' and
``classical'' behavior realized in many possible different materials. 

Last but not least, we consider realistic dipolar interactions which
either  from intermediary itinerant electrons,\cite{LiOsO3} 
from Coulomb forces,\cite{Devonshire-1949}
or by other effective mechanism, are expected to act as long range forces. 
Even when truncated at second neighbors, long range dipole-dipole interactions
give rise to new phases in a richer dipole-elastic phase diagram. 
The main results will confirm, as in the minimal pantograph model, the emergence of a spontaneous bulk polarization at zero magnetic field, 
as well as a sharp drop thereof once the magnetic field exceeds a critical value.


\subsection{Next-to-nearest neighbors and easy axis interactions}


\subsubsection{Electro-elastic sector extensions \label{sec: Electro-elastic sector extensions} }

The pantograph model in discussion is partly inspired in the material AgCrS$_2$ \cite{Streltsov}.
To start this Section we cast now some salient features, from the experimental understanding of this material, that motivate our modeling.

In  AgCrS$_2$ the magnetic ions Cr$^{3+}$ are arranged in triangular layers, each one surrounded by six S$^{2-}$ non-magnetic sulfur ions on 
the vertices of non regular octahedra (defining non equivalent crystallographic positions 
breaking the reflection symmetry with respect to the Cr plane).
It suffers a transition from the paramagnetic  $R3m$ structure
to a magnetically ordered phase with non centro-symmetric $Cm$ structure.  \cite{Damay-2011} 
The low temperature magnetic order is given by parallel ferromagnetic lines along one of the triangular layer axes, which alternate with the $\uparrow\uparrow\downarrow\downarrow$ pattern in the transverse direction.
This transition produces a magnetostriction enlarging (shortening) the distance between parallel (antiparallel) magnetic moments \cite{Streltsov}, 
then producing a shift of the center of charge of surrounding sulfur ions and a consequent spontaneous polarization.
As all the octahedra along a ferromagnetic line suffer the same distortions, 
the active elastic degrees of freedom can be effectively described by a one dimensional model   
across the $\uparrow\uparrow\downarrow\downarrow$ order (see Fig. (3) in [\onlinecite{Streltsov}]).
This motivates the parametrization of dipolar moments as in Eq.\ (\ref{eq: beta}), 
as well as the dipole-dipole coupling in Eq.\ (\ref{eq: H dip full}), used all along Section \ref{sec: Pantograph I} 
in combination with a simpler magnetic model. 

Moreover, once established that NNN  interactions play a central role in the magnetic sector, 
we also propose to consider an expansion of the dipolar interactions in Eq.\ (\ref{eq: H dip full}) up to second neighbors.
We expect that this inclusion could bring into play enough frustration in the antiferroelectric order, 
such as to change the phase diagram in Fig.\ \ref{fig: minimal electro-elastic}.
On the other hand, we also expect that the inclusion of third- and longer range terms will not modify qualitatively the arising dipolar phases,
at least for bipartite lattices where further neighbors fall into 
either the first or the second neighbor sublattices and will only renormalize the frustration.
Assuming small deformations as before,  we expand the coupling dependence on distance up to linear terms in distortions. 
We get 
\begin{widetext}
\begin{eqnarray}
H_{\text{dipole}}^{NNN}&=&J_{e}\sum_{i}\left(\sigma_{i}\sigma_{i+1}+\frac{1}{8}\sigma_{i}\sigma_{i+2}\right)
-2\varepsilon\sum_{i}\sigma_{i}
+2\beta\varepsilon\sum_{i}\delta_{i}\sigma_{i} \label{eq: H-dip}  \label{eq: H dip NNN} \\
&-&J_{e}\sum_{i}\left[
\left(\beta+\frac{3}{2a}\right)\left(\sigma_{i-1}\sigma_{i}+\sigma_{i}\sigma_{i+1}\right)
+\frac{1}{8}\left(\beta+\frac{3}{4a}\right)\left(\sigma_{i-2}\sigma_{i}+\sigma_{i}\sigma_{i+2}\right)
+\frac{3}{16a}\sigma_{i-1}\sigma_{i+1}
\right]\delta_{i}, \nonumber
\end{eqnarray}
\end{widetext}
where $J_{e}$ and $\epsilon$ have been defined in Eq.\ (\ref{H dip minimal}).  
Though this expression may look cumbersome, it is quadratic in dipolar variables $\sigma_i$ 
coupled by linear interaction vertices to elastic distortions. 

\begin{figure}[ht]
	\begin{centering}
		\includegraphics[scale=0.7]{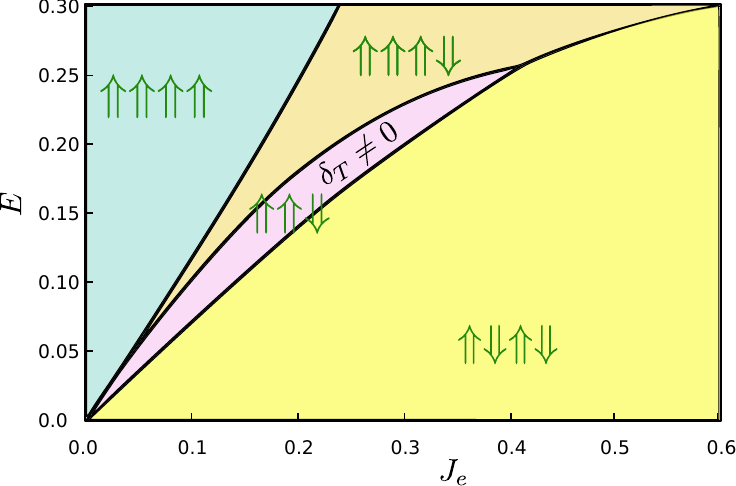} 
		\par\end{centering}
	\caption{Electro-elastic phase diagram (scale corresponds to original data for $\beta=0.2$, $K=1$ in Ref.\ [\onlinecite{pantograph-1}]). Elastic distortions follow the dipole pattern periodicity, except in the zero field line $\epsilon=0$ and the saturation region $\Uparrow\Uparrow\Uparrow\Uparrow$ where magnetic ions are equally spaced.  
	 A region with period three distortions $\delta_T$ (trimerized) stabilizes because of long range interactions.
	 }
	\label{fig: extended electroelastic} 
\end{figure}

Let us discuss the polarization due to an external electric field stemming from Eq.\ (\ref{eq: H dip NNN}), 
that is the  pantograph model 
when the magnetic sector is decoupled from the classical degrees of freedom ($\alpha=0$).
To this end we analyze the minimum energy configurations of the dipole-Peierls Hamiltonian $H_{\text{dipole}}+H_{\text{elastic}}$: 
given different periodic dipolar patterns we analytically compute
the distortions minimizing the elastic energy, in the presence of the electric field.
By comparison we select the lowest energy electro-elastic configuration.
In detail, we have considered all of the ordered dipolar configurations up to period four. 
The results lead to the dipole-elastic phase diagram in Fig.\ \ref{fig: extended electroelastic}. 

Without electric field the system possesses a $\mathbb{Z}_2$ inversion symmetry, but spontaneously adopts one of the two possible antiferroelectric $\Uparrow\Downarrow\Uparrow\Downarrow$ configurations. 
The distortions are null in either configuration, then dipoles pointing up or down have the same magnitude and the system  has no net polarization.

When a small electric field is turned on, breaking the inversion symmetry, 
no dipole flips are produced below a critical field but alternate distortions are induced
\begin{equation}
	\delta_i=-(1)^{i} \frac{p_0 \beta}{K} \epsilon,
	\label{eq: delta AFE}
\end{equation}
meaning that bonds get shorter where dipoles point along the field, 
enlarging the corresponding local dipolar momenta 
while lowering the  strength of dipoles pointing in the opposite direction.
Then the system behaves as a simple paraelectric,
acquiring a bulk polarization proportional to the applied electric field (with electric susceptibility $\chi_e=\frac{\partial P}{\partial E}=\frac{2 p_0^4 \beta^2}{K})$. 

At the critical line that separates the antiferroelectric low field phase from longer period dipolar structures, 
polarization gets discontinuous because of extensive dipolar flips.


\subsubsection{Magneto-elastic sector extensions}

As in the minimal model, the magnetic ions interact via super-exchange couplings.
Along with NN antiferromagnetic  couplings $J_{1}$, frustration is introduced 
by NNN antiferromagnetic couplings $J_{2}$.
The role of NNN couplings is to favor antiparallel spins every two sites,
so that a highly frustrated regime with $J_2/J_1$ above a critical value 
will be responsible for the $\uparrow\uparrow\downarrow\downarrow$ magnetic order.

The magneto-elastic model, both NN and NNN super-exchange couplings may depend on elastic distortions. 
However, we assume for simplicity that only the NN exchange $J_{1}$ shows a linear dependence 
on distortions, that written as  in Eq.\ (\ref{eq: J1_i}) reads
$
J_1(\delta_i)=J_1 (1-\alpha \delta_i)
$
while $J_2$ is not altered.
To support this assumption on the invariance of $J_2$, 
notice that in the frequent case of alternating distortions the 
second neighbor distances $2a+\delta_i+\delta_{i+1}$ are not altered at all. 

The effect of crystal fields, describing interactions with the magnetic ions environment,
leads in general to anisotropic spin interactions. 
Assuming that a local preferred direction exists, we introduce axially symmetric interactions:
the $SU(2)$ invariant Heisenberg interaction $\mathbf{S}_{i}\cdot \mathbf{S}_{j}$ is replaced by 
\begin{equation}
	\left(\mathbf{S}_{i}\cdot \mathbf{S}_{j}	\right)_\gamma
	\equiv 
	S_{i}^{z}S_{j}^{z}+\gamma \left(S_{i}^{x}S_{j}^{x}+S_{i}^{y}S_{j}^{y}\right).
	\label{anisotropic spin product}
\end{equation}
($z$ axis determined by the crystal environment).
Here $\gamma$ is the axial anisotropy parameter; 
aiming to describe collinear multiferroic materials, we focus on $\gamma \leq 1$;
that is, we cover from the easy axis anisotropy case $\gamma \ll 1$ 
to the isotropic case $\gamma=1$.
This is motivated by the large variety of known multiferroic materials, 
but also by the theoretical importance of the $SU(2)$ invariant point case.
The easy plane regime $\gamma > 1$,  not discussed here, 
is known to be continuously connected with the isotropic case 
(see for instance [\onlinecite{Tsvelik-book}]; 
it is usual to find in the literature a parameter $\Delta\equiv 1/\gamma$ to describe the XXZ spin chain
as a perturbation of the planar XY case).
For our purpose, the limit $\gamma \to 0$ connects our work with 
the classical Ising regime.

The magnetic sector, coupled to lattice distortions,  is then described by the Hamiltonian 
%
\begin{eqnarray}
	H_\text{spin}^{NNN}&=&\sum_{i} J_1(\delta_i) \left(\mathbf{S}_{i}\cdot \mathbf{S}_{i+1}	\right)_\gamma
	+\sum_{i}J_{2}\left(\mathbf{S}_{i}\cdot \mathbf{S}_{i+2}	\right)_\gamma \nonumber\\
	&-&h\sum_i S^z_i.
	\label{eq: H spin NNN}
\end{eqnarray}
A model described just by $H_{\text{elastic}}+ H_{\text{spin}}^{NNN}$ might be called a frustrated anisotropic spin-Peierls system.


\paragraph{Purely magnetic sector.}

In the absence of deformations the magnetic model in Eq.\ (\ref{eq: H spin NNN}) has been thoroughly studied.
We do not intend to cover the subject in all details but summarize the main results relevant 
for the present work;
for a complete treatment with a careful account of the literature see [\onlinecite{Giamarchi}] 
and references therein.

\begin{figure}[ht]
	\begin{centering}
		\includegraphics[scale=0.7]{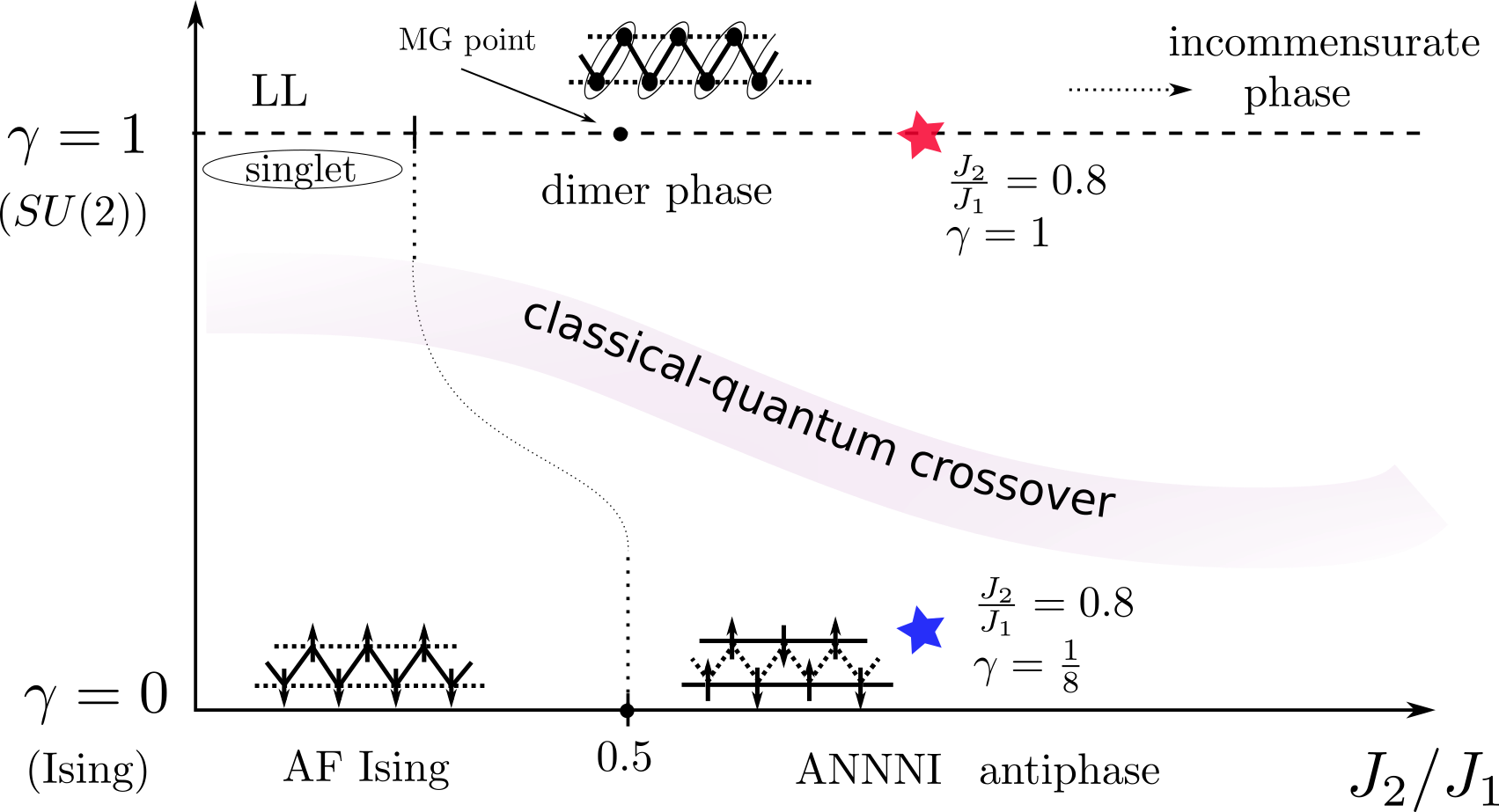}
		\par
	\end{centering}
	\caption{
		Schematic ground state diagram for the spin $S=1/2$ anisotropic frustrated antiferromagnetic chain. 
		Zig-zag miniatures of the spin chain are used to visually emphasize the prevalence of first neighbors or  second neighbors antiferromagnetic correlations, or singlet correlations indicated by ellipses.  A diffuse line separates the quantum behavior at low anisotropy from the classical behavior at high anisotropy.
		Most of the materials we are interested in are located in the frustrated anisotropic region
		(low right corner).  
		As representative points we numerically explore in detail a highly frustrated regime given by  $J_2/J_1=0.8$, 
		in the isotropic case  $\gamma=1$  (red star) and a high easy axis anisotropy $\gamma=1/8$  (blue star). 
	}
	\label{fig: frustration-anisotropy} 
\end{figure}

For our purpose the effects of frustration  $J_2/J_1$ and anisotropy $\gamma$ should be recalled.
When  no magnetic field is turned on
the several scenarios
can be summarized by the diagram in Fig. \ref{fig: frustration-anisotropy}. 

The anisotropy parameter $\gamma<1$   weakens the
quantum fluctuations of the transverse spin components, making the
spins ``more classical''. 
For systems with collinear order the zero $\gamma$ limit is equivalent to considering large $S$ spins,
in the sense that in a Holstein-Primakov \cite{Auerbach} expansion transverse fluctuations are suppressed out by a $1/S$ factor. 
Other approaches describe the easy axis component with a strong single ion anisotropy
\cite{Seno-1994}, or do instead introduce quantum fluctuations on top of classical spins
\cite{Dyson-Maleev-1956,Blanco-2017}.

For low frustration $J_2 \ll J_1$ the system can be seen as a linear antiferromagnetic chain $J_1$ weakly perturbed by NNN interactions $J_2$; 
in the opposite limit $J_2 \gg J_1$ it is better described as two-leg ladder of linear antiferromagnetic chains $J_2$ weakly coupled by zig-zag rungs $J_1$.
The $SU(2)$ symmetric line $\gamma=1$ is well studied by many techniques, 
in particular the bosonization of the effective low energy excitations  \cite{Haldane-1982}: 
for low frustration the ground state is a gapless Luttinger Liquid (LL) with quasi long range order
but enters a two-fold degenerate gapped quantum dimer phase for
$J_2/J_1>0.2411$ \cite{Okamoto-1992,Eggert-1996}, 
with expectation value of the local spin $\langle S^z_i\rangle =0$ and   
strong  antiferromagnetic (negative) spin correlations every two-bonds
(strictly, this is not collinear).
A paradigmatic example is found at $J_2/J_1=0.5$, the Majumdar-Ghosh point \cite{Majumdar-Ghosh-1969}, 
where the exact ground state is a (two-fold degenerate) direct product of two-site spin singlets. 
For very large frustration the gap decreases exponentially and the ground state   
shows incommensurate spiral spin correlations \cite{White-Affleck-1996,Allen-Senechal-1997,Nersesyan-1998}.
On the bottom of the diagram, the large anisotropy limit $\gamma = 0$ defines the one dimensional antiferromagnetic Anisotropic Next Nearest Neighbors Ising (ANNNI) model; 
classical spins order in a two-fold degenerate $\uparrow\downarrow\uparrow\downarrow$  Néel phase for low frustration ($J_{2}/J_{1}<0.5$) 
with a transition to the $\uparrow\uparrow\downarrow\downarrow$ antiphase state for larger frustration ($J_{2}/J_{1}>0.5$) \cite{Selke-1988}. 
In a sense, while $\gamma \to 0$  the LL quantum phase evolves into the classical Néel phase 
and the quantum dimer phase evolves into the $\uparrow\uparrow\downarrow\downarrow$ classical phase. 
Many of the materials we are interested in are located in the frustrated, easy axis anisotropic region 
(low right corner).  
Others correspond to the frustrated  ANNNI model with ferromagnetic $J_1<0$ and antiferromagnetic $J_2>0$, leading to the same  $\uparrow\uparrow\downarrow\downarrow$ antiphase state when $J_{2}/|J_{1}|>0.5$.


\paragraph{Magneto-elastic sector.}

When the magnetic sector is coupled to the lattice through $\alpha \neq 0$, 
the ground state magnetic configuration comes along with lattice distortions.
In the absence of dipolar degrees of freedom 
this interplay between distortions and modulated exchange couplings is resolved as 
an energy balance between elastic cost and magnetic energy gain. 
Technically, this balance is expressed by self consistent equations included in Eq.\ (\ref{eq: self-consistency-full}). 
The spin-Peierls mechanism that promotes the
formation of spin singlets at the cost of dimerized distortions in the NN antiferromagnetic chain \cite{Cross-Fisher-1979,Bray-1982},
has been proved to work also in the frustrated isotropic case \cite{Feiguin-etal-1997}. 
In general, when non trivial distortions show up in the ground state, the spin excitation spectrum is gapped. 
In consequence the magnetization curve presents a plateau:  a finite magnetic field is required for the Zeeman energy
to overcome the energy gap and change the spin state.
The spin-lattice coupling also
provides mechanisms for the opening of plateaus at different magnetization
fractions, either for quantum $S=1/2$ spins \cite{Cabra-Stauffer-2006}
or classical spins. \cite{Vekua-etal-2006}

Related to these magnetic features, notice that a  period three pattern
$\Uparrow\Uparrow\Downarrow$  
shows up in the electro-elastic sector of the extended model 
(see Fig.\ \ref{fig: extended electroelastic}), 
which is not captured when dipolar interactions are truncated at nearest neighbors (cf. Fig.\ \ref{fig: minimal electro-elastic}).
In this regime distortions occur with the same periodicity three and will eventually contribute or interfere with 
the period three $M=1/3$ magnetic plateau state that is expected for the 
magneto-elastic sector. \cite{Tertions}
We defer to Section \ref{sec: double frustration} the  analysis of the complete pantograph model 
at simultaneous fractional polarization and magnetization.


\subsubsection{Extended magneto-electro-elastic  model and self-consistent equations.}

We are now in position to discuss the complete Hamiltonian for the extended pantograph model.
It reads
\begin{equation}
	H_{MEE}^\text{(extended)}=H_\text{elastic}+H_\text{spin}^{NNN}+ H_\text{dipole}^{NNN},
	\label{eq: H-NNN}
\end{equation}
with the explicit forms of $H_\text{elastic}$, $H_\text{spin}^{NNN}$ and $H_\text{dipole}^{NNN}$ 
given in Eqs.\ (\ref{eq: H elastic}), (\ref{eq: H spin NNN}) and (\ref{eq: H dip NNN}) respectively.

As before in Section \ref{sec: minimal}, we follow the self-consistent approach to this Hamiltonian,
For a given configuration of dipoles $\sigma_{i}$ and a (quantum or classical) state for the spins $\mathbf{S}_{i}$, 
the minimal elastic energy is obtained when distortions $\delta_{i}$ satisfy the local zero gradient conditions
\begin{widetext}
\begin{eqnarray}
\label{eq: self-consistency-full}
K\delta_{i} &=& \alpha J_{1}\langle S_{i}^{z}S_{i+1}^{z}+\gamma\left(S_{i}^{x}S_{i+1}^{x}+S_{i}^{y}S_{i+1}^{y}\right)\rangle-\beta\varepsilon\sigma_{i}\\
&+&J_{e}\left(\beta+\frac{3}{2a}\right)\left(\sigma_{i-1}\sigma_{i}+\sigma_{i}\sigma_{i+1}\right)
+\frac{1}{8}J_{e}\left(\beta+\frac{3}{4a}\right)\left(\sigma_{i-2}\sigma_{i}+\sigma_{i}\sigma_{i+2}\right)+
J_{e}\frac{3}{16a}\sigma_{i-1}\sigma_{i+1},\nonumber
\end{eqnarray}

\end{widetext}
further constrained by the fixed chain length condition.
In comparison with the minimal model, Eq.\ (\ref{eq: self-consistency}), 
the novelties here are the NNN dipolar correlations and the relative importance of the easy axis spin correlations.

We recall that, on the one hand, these self-consistent equations allow to explore the interplay 
between magnetic and electric degrees of freedom either collaborating or competing to produce the optimal elastic distortions. 
Each of them enters in the form of local correlations.
On the other hand it allows to incorporate the knowledge about the magnetic sector and the electric sector separately. 
For the spin sector we take input below  both 
from known theoretical frameworks and from DMRG numerical computations.
In the present pantograph model NNN magnetic interactions are not explicit in Eq.\ (\ref{eq: self-consistency-full}).  
Nonetheless they play a central role in the actual value of the explicit NN correlations 
by introducing magnetic frustration in the Hamiltonian in Eq.\ (\ref{eq: H spin NNN}). 

The various parameters in the complete model (\ref{eq: H-NNN}) allow for a rich phase diagram. 
According to the multiferroic materials we aim to describe,
the main region of interest along the present work will be that with
large enough ratio $J_{2}/J_{1}$ so as to manifest magnetic frustration. 
For large anisotropy $\gamma \ll 1$ one could expect that spin fluctuations are strongly
diminished, allowing for a 
$\uparrow \uparrow \downarrow \downarrow$
ground state comparable  to the classical ANNNI model antiphase state. 
Fig.\ \ref{fig: frustration-anisotropy}, drawn for a purely magnetic model, serves as a reference.

The dipolar exchange $J_{e}$ will be kept below the magnetic exchange couplings, 
so that in principle it is magnetism what drives electric responses. 
The lattice stiffness $K$ will set an energy scale larger than magnetic and electric
ones, in order to keep distortions small with respect to the lattice
spacing $a$.
We set the length scale by taking the lattice spacing $a=1$ and also set the
energy scale taking $Ka^{2}=1$. 

From the above considerations, we choose for numerical computations a reference set of phenomenological parameters 
$J_{1}=0.5$, $J_{2}=0.4$ and $J_{e}=0.2$ to organize the energy scale of each degree of freedom.
We also choose $\alpha=\beta=0.2$ to analyze the magneto-elastic
and electroelastic couplings. Notice that our
results do not depend on fine tuning, so we expect them to be valid
in a wide region of parameters. 

Finally, the magnetic anisotropy will
be varied from  the quantum SU(2) symmetric point $\gamma=1$ down to small enough values to explore the large easy axis anisotropy regime 
where classical behavior is expected.


\subsection{Magnetization response  of the extended pantograph model}

In the complete model (\ref{eq: H-NNN}) the magnetic frustration ($J_1-J_2$) and the magneto-elastic mechanism ($\alpha\neq 0$)  co-exist, 
complemented with a dipolar energy cost/gain associated to lattice distortions. 
Altogether, this is expressed in the complete self consistent Eqs.\ (\ref{eq: self-consistency-full}).
These SC equations show that the pantograph mechanism puts dipolar and magnetic correlations
in either cooperation or competition with each other to produce changes in the bond lengths.
As in the minimal model, this is the key ingredient that provides an effective magneto-electric coupling mediated by lattice distortions.

We organize our analysis by we first building  magnetization curves  at zero electric field, 
addressing in particular  to the existence of  magnetization plateaus.
We focus on the region with high enough frustration so as to produce the $\uparrow\uparrow\downarrow\downarrow$ 
magnetic ordering;
for numerical work we take as a representative case the parameters $J_1=0.5$. $J_2=0.4$, $J_e=0.2$, $\alpha=\beta=0.2$.
Along the anisotropy range $\gamma \leq 1$  we find qualitatively different behaviors;
we report as representative examples 
the $SU(2)$ symmetric case $\gamma=1$ and a highly anisotropic case $\gamma=1/8$ 
(see red and blue stars, respectively, in  Fig. \ref{fig: frustration-anisotropy}).

We solve the self-consistent equations iteratively, 
feeding in the spin-spin correlations computed by DMRG in the presence of distortions 
and the zero electric field antiferroelectric dipolar configuration (see Fig. \ref{fig: extended electroelastic}). 
It is worth noticing that in this regime the dipolar degrees of freedom $\sigma_i$ are not excited. 
Once this is known, the full model in Eq.\ (\ref{eq: H-NNN}), 
can be seen as a frustrated magneto-elastic spin-Peierls Hamiltonian 
where distortions carry an extra energy cost due to the long range interaction
of antiferroelectrically ordered, distortion modulated, electric dipoles.

\begin{figure}[ht]
	\begin{centering}
		\includegraphics[scale=0.80]{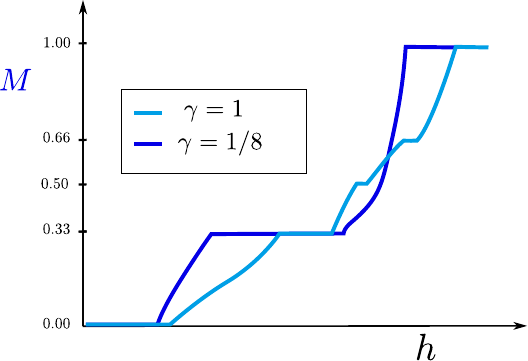} 
		\par\end{centering}
	\caption{
		Schematic magnetization curves obtained for the extended pantograph model, in the isotropic case ($\gamma=1$) and  
		 a highly easy axis anisotropic case ($\gamma=1/8$) (reproducing actual for $J_1=0.5$. $J_2=0.4$, $J_e=0.2$, $\alpha=\beta=0.2$ and zero electric field, see Ref.\ [\onlinecite{pantograph-2}]).
		A plateau at $M=0$ and a prominent plateau at magnetization fraction $M=1/3$ are observed in both cases; 
		higher magnetizations $M=1/2,2/3$ form narrow plateaus only in the isotropic case.
	}
	\label{fig: magnetization-curves} 
\end{figure}

By computing the energy of all the possible magnetizations in a finite size chain of length $N$ 
we draw the magnetization curves. 
Representative cases are shown in Fig.\ \ref{fig: magnetization-curves}
where the magnetization $M$ is defined as the total $\langle S^z_{\text{total}} \rangle$ relative to saturation.

The outcome is a very rich phase diagram that not only includes previously studied situations, 
but also suggests some exotic non-trivial ones.
Besides the $M=0$ plateau, present for both the isotropic and the anisotropic case, 
one can see other plateaus at simple fractions of the saturation magnetization.
In particular, there is a noticeable plateau at $M=1/3$ that is much wider in the anisotropic case, 
and comes together with a period three distortion modulation.
There are also  plateaus at $M=1/2$ and $M=2/3$ in the isotropic case, 
which are no longer present in the anisotropic case for $\gamma = 1/8$. 

\begin{figure}[ht]
\begin{centering}
	\includegraphics[scale=0.7]{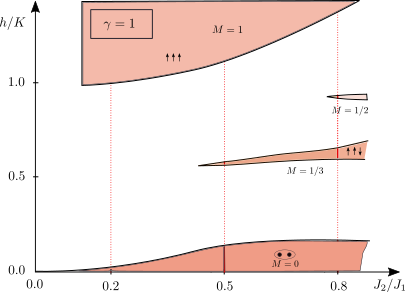}  
	\par
\end{centering}
\begin{centering}
	\includegraphics[scale=0.7]{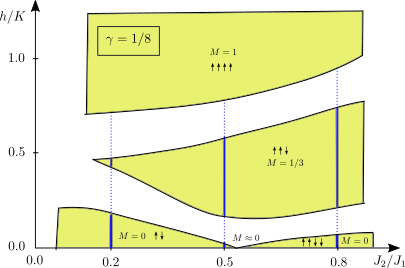}  
	\par
\end{centering}
\caption{Schematic magnetic phases showing plateaus from
	the competition between the frustrated exchange $J_{2}/J_{1}$ and
	the magnetic field $h$ in 
 	the isotropic case (top panel, $\gamma=1)$ and
	a highly anisotropic case (bottom panel, $\gamma=1/8)$. 
	Legends: $M$ is the magnetization relative to saturation, arrows indicate classical collinear order and points in an ellipse indicate quantum singlet dimers. 
	The vertical lines correspond computed to magnetization curves at $J_{2}/J_{1}=0.2,\, 0.5 \text{ and } 0.8$, 
	($J_1=0.5$, $J_e=0.2$, $\alpha=\beta=0.2$).}
\label{fig: H vs J2/J1} 
\end{figure}

For completeness, we have also computed the magnetization curves for systems with some lower frustration values ($J_2/J_1=0.2,\,0.5$).
We sketch in Fig.\ \ref{fig: H vs J2/J1} a graphical 
summary of the observed plateaus according to the applied field $h$ and the magnetic frustration measured by $J_2/J_1$ 
(the low frustration area at the left is not depicted since it has been thoroughly studied in the literature and is not relevant for our purposes).

\subsubsection{Zero magnetization plateau}

In this Section we discuss the zero magnetization plateau and the magnetic excitations that make the system exit from it,
with emphasis on the description of experimental setups attainable in the multiferroic materials
surveyed in the Introduction. 
Also, for theoretical interest we compare the magnetic structure 
of the $M=0$ plateau state observed in the $SU(2)$ isotropic case ($\gamma=1$) 
and the easy axis anisotropic case ($\gamma=1/8$).  
In spite of their differences, we will show that both of them lead to alternating distortions and 
produce a finite bulk polarization  at zero electric field.
Quantum fluctuations, though substantially damped, turn out to be relevant 
even for the (Ising-like) large anisotropic limit.

\vspace{2mm}

\paragraph{Quantum dimerized plateau} $ $
\vspace{2mm}

We recall that, without exchange modulation ($\alpha=0$),   
the homogeneous isotropic ($\gamma=1$) frustrated spin $S=1/2$ antiferromagnetic Heisenberg chain 
spontaneously breaks the translation symmetry and enters a quantum dimer phase for $J_2/J_1>0.2411$,
\cite{Okamoto-1992,Eggert-1996,Giamarchi}
with $\langle S^z_i\rangle =0$ and spin correlations dominated by strong antiferromagnetic (negative) values every two-bonds. 
In the limiting case $\langle \mathbf{S}_i\cdot \mathbf{S}_{i+1} \rangle =-3/4$ one would find two-spin singlets \cite{Majumdar-Ghosh-1969}, 
while correlations close to such limit are called spin dimers.
These dimers can form in even or odd bonds, making the ground state two-fold degenerate.

In the presence of the magneto-elastic coupling in Eq.\ (\ref{eq: J1_i}) the NN spin-spin correlations 
have influence on elastic distortions, as seen in the first line of Eqs.\ (\ref{eq: self-consistency-full}).
As the frustrated spin-spin correlations alternate along the chain, 
frustration favors alternating distortions with spin singlets located at short bonds. 
Regarding the electro-elastic coupling, one can see that the antiferroelectric configuration at zero electric field
has site independent dipole-dipole correlations (negative between first neighbors, positive between second neighbors). 
According to the second line in  Eqs.\ (\ref{eq: self-consistency-full}), and taking into account the fixed length constraint,  
dipole-dipole correlations have no influence on distortions. 
However, the strength of the dipoles is influenced by distortions. Dipoles sitting in shortened bonds are enlarged, 
while those sitting in enlarged bonds are shortened (see Eq.\ (\ref{eq: beta})).
Here, as in the minimal pantograph model,  the magnetic frustration gives rise to a ferrielectric state, 
carrying a spontaneous bulk electric polarization.

Such a bulk ferrielectric polarization
has been observed in several multiferroic materials.
In particular we mention again the case of  AgCrS$_2$, \cite{Streltsov}
with a crystal structure closely related to
delafossites. 
In this material the magnetostriction is manifest in a quasi one dimensional setting
directly comparable  with the present extended pantograph model.

The present analysis for the frustrated isotropic magneto-elastic chain 
makes more robust our results in Section \ref{sec: minimal} for the minimal pantograph model,
where in the absence of frustration 
spontaneous polarization is only due to the spin-Peierls instability of nearest neighbors Heisenberg spin chains.
\cite{pantograph-1}
\begin{figure}[ht]
	\begin{centering}
		\includegraphics[scale=0.4]{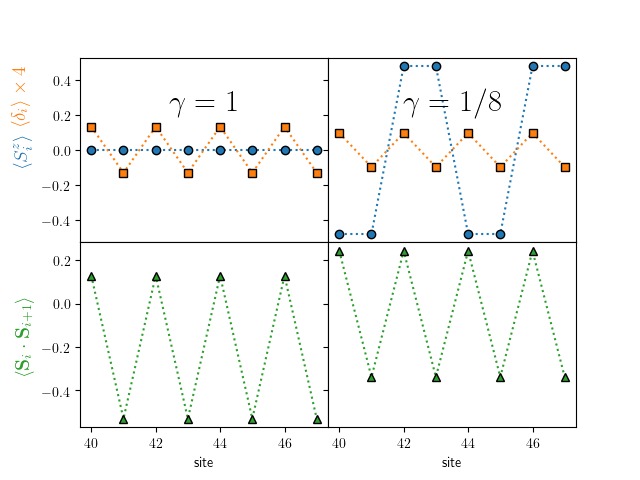}  
		\par\end{centering}
	\caption{Numerical results for the
		$M=0$ plateau configuration in a chain of 84 sites 
		with periodic boundary conditions ($J_1=0.5$. $J_2=0.4$, $J_e=0.2$, $\alpha=\beta=0.2$, 
		in the presence of an antiferroelectric dipolar background; reprinted from Ref.\ [\onlinecite{pantograph-2}]).
		Upper panels: local profile of $\langle S^z _i\rangle$ (blue circles) and distortions $\delta_i$ (orange squares), 
		in the isotropic case (left panels, $\gamma=1$) and highly anisotropic case (right panels, $\gamma=1/8$). Distortions are scaled by a convenient factor for better visualization.
		Lower panels: local profile of spin correlations $\langle \mathbf{S}_i\cdot  \mathbf{S}_{i+1}\rangle$ in the isotropic and anisotropic cases.
	}
	\label{fig: M=0 profiles} 
\end{figure}
Numerical support is shown in the left panels of Fig.\ \ref{fig: M=0 profiles} the local spin expectation value,  
the distortion profile and spin-spin correlations obtained by solving 
Eqs.\ (\ref{eq: self-consistency-full})  for $J_1=0.5$. $J_2=0.4$, $J_e=0.2$, $\alpha=\beta=0.2$, $\gamma=1$.
A cartoon description is shown in Fig. \ref{fig: polarized quantum dimers} to help reading the data plots.
\begin{figure}[ht] 
	\begin{centering}
		\includegraphics[scale=0.55]{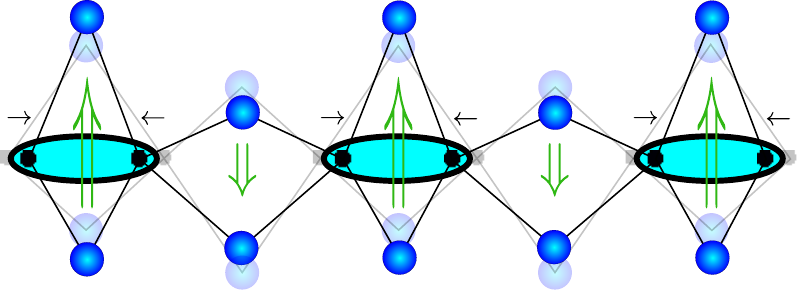} 
		\par\end{centering}
	\caption{ Schematic picture for the quantum plateau state at $M=0$. 
		The dimer singlets represented by ellipses gain magnetic energy by shortening their distance, thus enlarging their exchange coupling. The influence of these distortions on the alternating dipoles lengths (green double arrows) produces a ferrielectric configuration with a finite bulk polarization.
	}
	\label{fig: polarized quantum dimers} 
\end{figure}

\vspace{2mm}

\paragraph{Classical $\uparrow\uparrow\downarrow\downarrow$ plateau} $ $
\vspace{2mm}

In the easy axis anisotropy limit $\gamma \to 0$ and no magneto-elastic coupling  ($\alpha=0$) 
our model becomes  the homogeneous frustrated antiferromagnetic Ising chain (ANNNI model). 
We recall that this model enters the collinear state
$\uparrow\uparrow\downarrow\downarrow$ (called antiphase in that context) at $J_2/J_1>0.5$, \cite{Selke-1988}
because $J_2$ is large enough to make the NNN spin correlations everywhere antiferromagnetic, 
while NN correlations alternate between FM and AFM with $\pm S^2$.
Same as in the quantum case, the analysis of the self-consistent conditions in Eq.\ (\ref{eq: self-consistency-full}) 
shows that the magneto-elastic terms favor alternating distortions, inducing the $\mathbb{Z}_2$-symmetric
spontaneous ferrielectric polarization.

\begin{figure}[ht]
	\begin{centering}
		\includegraphics[scale=0.55]{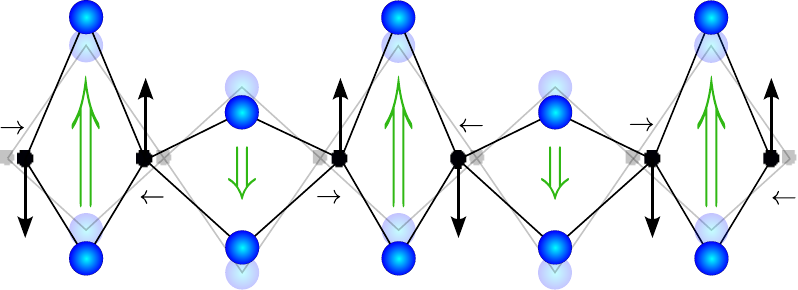} 
		\par\end{centering}
	\caption{Schematic picture for the classical $\uparrow\uparrow\downarrow\downarrow$ plateau state at $M=0$. 
		The collinear spin configuration  represented by black arrows gains magnetic energy by enlarging 
		the  exchange coupling of anti-parallel nearest neighbors, shortening their distance. 
		These distortions  produce a ferrielectric configuration (green double arrows) with a finite bulk polarization.
	}
	\label{fig: polarized uudd} 
\end{figure}

To explore this scenario we performed the DMRG self-consistent computation 
of the magnetic ground state for the same parameters as in the previous subsection, 
except for a markedly anisotropic easy axis spin-spin interaction, $\gamma=1/8$. 
We show in the right panels of Fig.\ \ref{fig: M=0 profiles} the spin and distortion profiles.
They indicate that the spins almost saturate the $z$ component, $\langle S^z_i \rangle \approx \pm 1/2$, following the $\uparrow\uparrow\downarrow\downarrow$ pattern. 
Spin-spin correlations are close to classical, with  $\langle \mathbf{S}_i\cdot  \mathbf{S}_{i+1}\rangle\approx 1/4$ for ferromagnetic bonds and $-1/4$ for antiferromagnetic bonds. 
The distortions do alternate, with short (long)  bonds when spin correlations are antiferromagnetic (ferromagnetic). 
Same as in the quantum dimerized plateau, alternating distortions lead to a finite spontaneous electric polarization.
A pictorial description of this state is shown in Fig.\ \ref{fig: polarized uudd}.

It is worth emphasizing the robustness of the spontaneous ferrielectric polarization induced by magnetic instabilities in the pantograph model.
We have found the same result in very different regimes, 
such as the magnetically frustrated $J_1-J_2$  quantum spin chain, the close to classical frustrated (Ising) chain, 
and the spin-Peierls chain without magnetic frustration. \cite{pantograph-1}
%
%

\subsubsection{Magnetic excitations}

The $M=0$ configuration remains stable under an external magnetic field $h$, 
until it reaches a critical value $h_c$ such that the gain in Zeeman energy of a magnetically excited state is larger than the spin gap. 
In this situation the system overpasses the $M=0$ plateau and enters a magnetized regime (see Fig.\ \ref{fig: magnetization-curves}). 
In order to understand the magnetization process we start by analyzing the features of the $S^z_\text{total}=1$ state; 
we then check that low magnetization states can be described as a superposition of elementary magnetic excitations.

\vspace{2mm}

\paragraph{Excitation of the quantum dimerized plateau} $ $
\vspace{2mm}

There exist extensive studies of the  $S^z_\text{total}=1$  excitation of the  $S=1/2$ magnetoelastic spin-Peierls Heisenberg chain,
which appears to be fractionalized into two spinons. \cite{Arai-1996}
In the bosonization framework 
these spinons can be explained as topological solitonic excitations 
of a sine-Gordon low energy effective continuum theory coupled to the distortion field. \cite{Fukuyama-1980}
Their presence has been checked numerically by different techniques \cite{Feiguin-etal-1997}
and they are found to condense at the ground state in the presence of a magnetic field.

Relevant to our purpose is the fact that the topological solitons connect different degenerate vacua of the system. 
In the spin-Peierls Heisenberg chain ($\gamma=1$) the ground state is two-fold degenerate and these vacua are the two possibilities 
of forming singlet pairs along the chain; that is, the two vacua differ by a one-site translation. 
The sequence of elastic distortions is also shifted by one site across each soliton, 
as the short bonds belong together with magnetic singlet pairs.

The self-consistent results \cite{pantograph-2} prove that solitons do develop in the present model,
when distortions are coupled to the amplitudes of antiferroelectrically ordered dipoles.
These solitons separate two different domains, say A and B, where the alternate distortion patterns are
displaced one lattice site with respect to each other. The magnetic excitation is fractionalized,  
with each soliton carrying $S^z=1/2$.
Spin-spin NN correlations alternate between strong antiferromagnetic ones (close to perfect singlets) 
and weakly correlated ones. 

\begin{figure}[ht]
	\begin{centering}
		\includegraphics[scale=0.6]{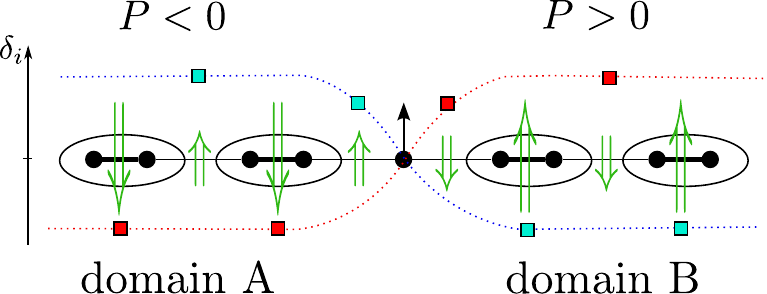} 
		\par\end{centering}
	\caption{
		Qualitative picture of a magnetic soliton connecting the two possible quantum dimer vacua.  
		Magnetic ions are represented by black circles.
		Bond distortions $\delta_i$ are represented by red (cyan) squares at odd/even sites. 
		Double arrows represent electric dipoles sitting amid magnetic sites, 
		in an antiferroelectric configuration.
		Dotted lines are a guide to follow the alternation of distortions (cf. similar actual data in Fig.\ \ref{fig: solitons E=0 minimal}).
		One unpaired magnetic site (with spin represented by a black arrow) carries the $S=1/2$ fractional magnetization of the soliton.  
	}
	\label{fig: site-shift} 
\end{figure}

A qualitative picture in Fig.\ \ref{fig: site-shift} illustrates the 
two different dimerized domains, A and B, separated by the soliton.
The bond distortions $\delta_i$  (drawn with squares with respect to a vertical axis) are alternate in each domain,
but red/cyan squares (say odd/even bonds) change sign; if odd bonds are short in domain A, they are long in domain B. 
Dimer singlets are signaled by ellipses, they are formed in short bonds within each domain
taking advantage of enhanced antiferromagnetic exchange.
One unpaired magnetic ion remains at the center of the soliton, pointing up with spin projection $S^z=1/2$. 
Within each domain the dipoles develop a ferrielectric net polarization, but
\textit{pointing in opposite directions} (dipoles represented by double arrows).

It is important that both domains are found to have approximately the same length. 
This is expected from the sine-Gordon low energy theory \cite{Manton_book}
and  numerically observed \cite{Mastrogiuseppe-2008} due to the exponential tails of the soliton profiles, 
which produce a residual repulsion between them. 
It has been also shown that for higher $S^z_\text{total}$ the excitations are pairs of solitons distributed as a periodic array, 
evolving into a sinusoidal magnetization profile.\cite{{Lorenz_1998}}  
This confirms the switch-off of electric polarization as magnetization grows and solitons proliferate.

\vspace{2mm}

\paragraph{Excitation of the classical $\uparrow\uparrow\downarrow\downarrow$ plateau} $ $
\vspace{2mm}

Given the Ising-like $\uparrow\uparrow\downarrow\downarrow$ structure found 
in the anisotropic case $\gamma=1/8$
for the $M=0$ plateau 
in Fig.\ \ref{fig: M=0 profiles} (top right panel), 
one could expect that the $S^z_\text{total} =1$ magnetic excitation also looks Ising-like, 
that is a simple spin flip followed by a rearrangement of classical spins defining sharp domain walls where some
second neighbors correlations get frustrated (ferromagnetic).

\begin{figure}[ht]
	\begin{centering}
		\includegraphics[scale=0.80]{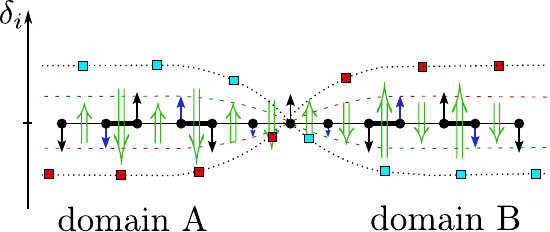}
		\par\end{centering}
	\caption{
		Schematic description of the soliton 
		connecting two different $\uparrow\uparrow\downarrow\downarrow$ dimerized domains. 
		Distortions of odd/even bonds are indicated with red/cyan squares in a vertical axis.
		One magnetic sublattice is not altered by the soliton (black arrows), 
		the other exhibits the reversal of spins (blue arrows) with projections following the soliton shape.
		Antiferroelectric dipoles (green double arrows) define
		 ferrielectric polarizations with opposite directions in each dimerized domain.
	}
	\label{fig: cartoon uudd} 
\end{figure}

However, it happens that the system takes advantage of quantum fluctuations to develop solitonic excitations, 
so that the reduction of $\langle S^z_i\rangle$ in the soliton region lowers the energy cost
of the frustrated second neighbors correlations.
We summarize the results in Ref.\ [\onlinecite{pantograph-2}] about these soliton features
with the visual aid  of Fig.\ \ref{fig: cartoon uudd}. 

Away from the soliton regions the alternation of distortions  
and the $\uparrow\uparrow\downarrow\downarrow$ spin pattern, 
are similar to the classical $S^z_\text{total}=0$ plateau structure but shifted 
by one lattice site across each soliton. 
However, a sublattice of magnetic ions every two sites (say odd sites, with spins represented by black arrows) 
keeps homogeneous $\uparrow\downarrow$ order across the soliton. 
Instead, the other sublattice (say even sites, with spins represented by blue arrows) 
exhibits the characteristic soliton shape, changing from $\uparrow\downarrow$ to $\downarrow\uparrow$
order in different domains. As a result, the soliton carries $S^z=1/2$ spin projection.

Regarding distortions, they follow the same pattern as in the isoptropic case (cf. Fig.\ 
\ref{fig: site-shift}). 
Then again  the electric dipoles form ferrielectric domains with the polarization
	pointing in opposite directions.

Notice that the solitons in the anisotropic case are slightly narrower than those in Fig.\ \ref{fig: quantum excited profiles}, 
for the isotropic case $\gamma=1$. 
The more anisotropic the interaction, one finds numerically that the soliton regions gets even narrower. 
But they do not evolve into sharp domain walls, 
at least for anisotropies as large as $\gamma = 0.01$.  
It is remarkable that quantum fluctuations play a significant role even in the quasi-classical limit.

The presence of topological solitons, instead of sharp domain walls, is decisive in the formation of equal length 
$\uparrow\uparrow\downarrow\downarrow$ domains: 
it is the repulsive residual interaction between solitons what keeps them separated in the finite size chain.


\subsection{Multiferroic macroscopic effects}


\subsubsection{Polarization jump driven by magnetic field}

At zero electric field, both in the isotropic and the anisotropic cases, 
the solitonic magnetic excitations separate ferrielectric domains with opposite polarization.  
This happens not only for $S^z_\text{total}=1$ but for higher excitations described by pairs of solitons.
As a consequence, having these domains the same length, the total polarization of the system drops nearly to zero. 
That is, the spontaneous electric polarization observed at zero magnetization is switched off 
by means of the applied magnetic field. \cite{pantograph-1}
This happens either if the exit from the $M=0$ plateau is smooth 
(that is, soliton pairs appear continuously with the magnetic field) 
or in the case of a metamagnetic jump in which soliton pairs proliferate.

\begin{figure}[ht]
	\begin{centering}
		\includegraphics[scale=0.4]{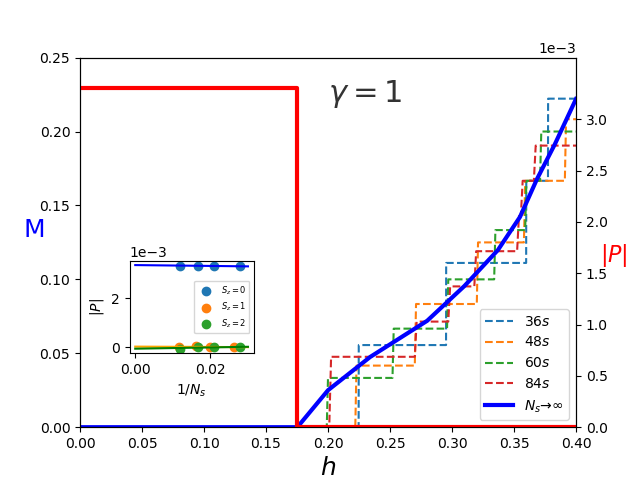}
		\includegraphics[scale=0.4]{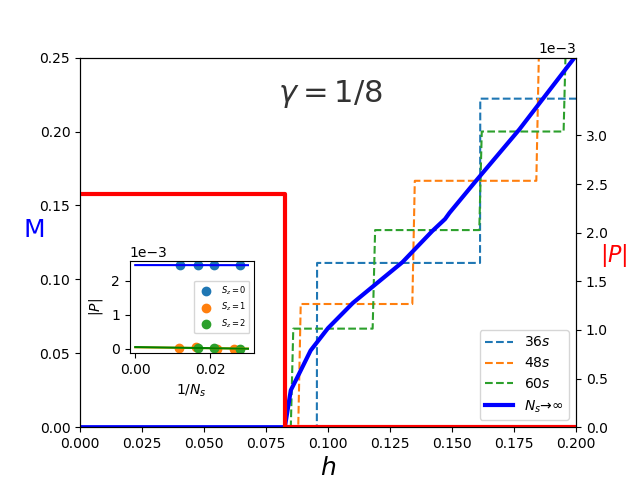} 
		\par\end{centering}
	\caption{
		Polarization curves (red solid lines, scale in the right axis in units of $p_0$) and 
		magnetization curves  
		(low $M$ region, finite size data and extrapolation as blue solid lines, scale in the left axis)
		in an external magnetic field 
		for the isotropic $\gamma=1$ and the anisotropic $\gamma=1/8$ models
		($J_1=0.5$, $J_2=0.4$, $J_e=0.2$, $\alpha=\beta=0.2$,
		in the presence of an antiferroelectric dipolar background; reprinted from Ref.\ [\onlinecite{pantograph-2}]).
		Insets: finite size scaling for the polarizations obtained for $S^z_\text{total}=0,\, 1,\, 2$ shows almost no size dependence. }
	\label{fig: polarization switch} 
\end{figure}

To make apparent the relation between the polarization jump and the onset of magnetization, 
we plot together in Fig.\ \ref{fig: polarization switch} 
the polarization and the low magnetization curves in a magnetic field, both for the isotropic (upper panel) 
and the anisotropic (lower panel) cases discussed along this work.
The spontaneous polarization (red curves, scale in right axis) is computed from the lattice distortions in an antiferroelectric background, 
according to Eq.\ (\ref{eq: polarization definition}). In both cases it suddenly drops several orders of magnitude.
The magnetization is the same as in Fig.\ \ref{fig: magnetization-curves}, with the addition of an infinite size extrapolation (blue curves, scale in left axis).
The infinite size extrapolation of the polarization at the lowest magnetization levels, shown in the insets, 
clearly proves that the polarization switch off is a bulk magnetoelectric effect occurring at the onset of magnetization. 
Beyond the excited $S^z_\text{total}=1$ and $S^z_\text{total}=2$ states, with polarization shown in the insets, 
we have checked that the further increase of the magnetization introduces extra pairs of solitons. 
These appear uniformly spread along the chain, as it also occurs in the magneto-elastic case, \cite{Lorenz_1998}]
separating different dimerization domains and producing the drop of the electric polarization observed 
in Fig.\ \ref{fig: polarization switch} for arbitrary non vanishing magnetization.

Such magnetically driven polarization jumps are a source of intrigue in many multiferroic materials. 
For instance, Lu$_2$MnCoO$_6$  \cite{Chikara-2016}
and  Er$_2$CoMnO$_6$ \cite{Oh-Oh-2019} show a polarization jump when exiting 
the observed $M=0$ magnetization plateau. 
Closely related are the polarization jumps observed in R$_2$V$_2$O$_7$ (R = Ni, Co)
when entering and exiting 
the $M=1/2$ magnetization plateau \cite{Ouyang-2018}. 
We expect that the present results could help in fitting actual parameters in these materials and explain the observed jumps.

\subsubsection{Polarization flip controlled by very low electric fields \label{sect: device}	}

Measures of spontaneous polarization are usually made with the help of a tiny poling field, 
to lift the degeneracy between the possible spontaneous orientations. 
Once done, a coercive field much larger than the poling one is required to flip the bulk polarization.

\begin{figure}
	\begin{centering}
		\includegraphics[scale=0.75]{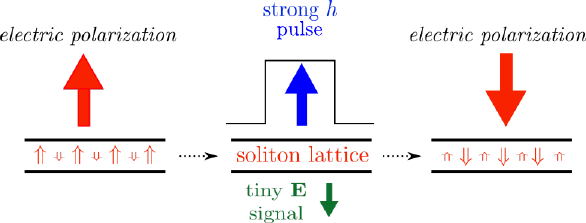}
		\par
	\end{centering}
	\caption{A magnetic field pulse, in any orientation and strong enough to magnetize the system, produces electric depolarization.
		In combination with a tiny poling field signal, it can be used to reverse the spontaneous polarization $\mathbf P$. 
		This could be the basis for storing information in a dipolar memory bit.
	}
	\label{fig: device} 
\end{figure}

In the present model it is also interesting to discuss the effect of a poling electric field
when the polarization has been switched off by a magnetic field larger than the critical one, 
strong enough to magnetize  and depolarize the system 
by the creation of pairs of different ferrielectric domains (see middle panel of Fig.\ \ref{fig: device}).
From this situation, as soon as the magnetic field is turned off, 
it is expected that the orientation of the  spontaneous polarization
follows the poling field direction.

\subsubsection{Magnetization jump driven by an electric field \label{subsec: extended M-vs-E}}

As discussed in  Section \ref{subsec: M-vs-P}, one can analyze the effects of
an external electric field on the magnetization curves.
The presence of the period four $\Uparrow \Uparrow\Uparrow\Downarrow$ dipolar configuration in the 
electro-elastic sector (see Fig.\ \ref{fig: extended electroelastic}) 
anticipates the opening of a spin gap and the stabilization of a magnetic plateau at $M=1/2$
(similar to that shown in Fig.\ \ref{fig: M vs H E=02 minimal electro-elastic}).
 
In full analogy with the scenario presented for the minimal pantograph model,
a variation of the electric field opening/closing  the magnetic spin gap 
would modify the magnetization curve and 
habilitate a magnetization jump driven by the electric field.
While no further differences deserve more detail in this case, 
the appearance of the period three  $\Uparrow\Uparrow\Downarrow$ configuration
provides an interesting new scenario, to be discussed in the next Section. 
%


\subsection{Extended pantograph model highlights}

We have shown here that in the extended pantograph model (\ref{eq: H-NNN}) the separate interplay of the spin and the dipolar sectors   
with the same lattice distortions gives rise to a spontaneous bulk polarization without the presence of an external electric field. 
In this context an external  magnetic field above a threshold marks the onset of magnetization simultaneously with a sharp  polarization switch-off. 
Thus the main achievements of the minimal pantograph model in Section \ref{sec: minimal} are  present in the more realistic extended model.
In addition, when a NNN coupling is strong enough we reproduce the experimentally observed
$\uparrow\uparrow\downarrow\downarrow$ spin pattern at zero magnetic field.

One can think of designing a multiferroic memory storage in which information, in the form of a polarized spot, 
is controlled by a low electric field signal with the help of a brief but strong magnetic blast: 
a magnetic field, carrying no information, would erase the previously "written" polarization, 
which is then "rewritten" in the desired (up or down) orientation 
by the simultaneous presence of a poling low electric field (low voltage bias). 
The procedure is sketched in Fig.\ \ref{fig: device}.
Such a device would show a giant electric response, and could be the basis for an efficient memory writing/reading device.


\section{Extended model at finite magnetization and finite polarization: double frustration \label{sec: double frustration} }

A salient feature of the electro-elastic sector in Section  \ref{sec: Electro-elastic sector extensions}
is 
the observation of an ordered dipolar phase with period three, which shows up in the presence of an appropriate homogeneous
external electric field (see Figure \ref{fig: extended electroelastic})
because of the long range character of dipolar interactions. We stress that this phase is not present in the minimal pantograph model 
where only nearest neighbors dipolar interactions are considered
(see Figure \ref{fig: minimal electro-elastic}).
Some properties of this dipolar phase, denoted as $\Uparrow\Uparrow\Downarrow$ in the following 
motivate this separate Section.

As before, we are interested on a parameter
region where the magnetic and dipolar couplings are of the same order
of magnitude, so both the spin and dipole configurations are relevant
to determine the ground state of the system. Also the magneto-elastic
coupling $\alpha$ and the electro-elastic coupling $\beta$ are similar,
in order to provide an efficient elastically mediated magneto-electric
interaction. We then reduce the free parameters in the Hamiltonian
(\ref{eq: H-NNN}) by taking $Ka^{2}$ as the energy unit
and fixing $J_{1}$, $J_{e}$, $\alpha$ and $\beta$ at convenient
values detailed below. Only $J_{2}$ and $\gamma\leq 1$ will be varied
to explore the incidence of magnetic frustration  and easy-axis anisotropy in the
ground state properties of the magnetized system. Different values of $J_{1}$,
$J_{e}$, and $\beta$ can be studied similarly in order to describe different materials.

External electric and magnetic fields $\boldsymbol{E}$ and $\boldsymbol{B}$
will be adjusted to drive the system to the peculiar double frustration
scenario we discuss here. This is the region where
the electric field polarizes the otherwise antiferroelectric dipolar
sector (driven by $J_{e}$) up to $P=1/3$ of saturation, provoking
the period three $\Uparrow\Uparrow\Downarrow$ dipolar pattern 
and the magnetic field sets
the spin degrees of freedom in the $M=1/3$ plateau region 
(see Figure \ref{fig: magnetization-curves}). 
For a magneto-elastic chain (not coupled to electric dipoles), this plateau is known to
appear together with an energetically favorable period three elastic
distortions. \cite{Vekua-etal-2006,Tertions,2007-Rosales-et-me}
On the other hand, for the electro-elastic chain obtained from the
Hamiltonian (\ref{eq: H-NNN}) when the spin sector is decoupled
($\alpha=0$), the $\Uparrow\Uparrow\Downarrow$ dipolar pattern also
comes along with period three elastic distortions  
bringing closer (farther) antiparallel
(parallel) dipoles. 

The question arises whether the elastic distortions compete or collaborate
in lowering the ground state energy of the magneto-electro-elastic
multiferroic system.
We then investigate the commensurability interplay of the $P=1/3$
period three dipolar order $\Uparrow\Uparrow\Downarrow$
with the period three magnetic
configurations observed in many frustrated magnetic materials with
$M=1/3$ magnetization plateaus .
In most magneto-elastic studies the $M=1/3$ plateau state is found to form a collinear
$\uparrow\uparrow\downarrow$ 
classical pattern \cite{Okunishi-2003}, but a {\em quantum  order}
$\bullet\!\!-\!\!\bullet\uparrow$ (where $\bullet\!\!-\!\!\bullet$
stands for a spin singlet dimer)
has been also predicted for spin $S=1/2$ modulated isotropic Heisenberg
chains \cite{Hida-Affleck-2005}. 
The robustness of magnetic
plateau states, given by a wide energy gap in the magnetization spectrum,
makes them good candidates for technological applications.

We describe here that they do compete, with profound
consequences  in the magnetic plateau configuration.
%


\subsection{Qualitative description of the double frustration scenario}

The self-consistent conditions in Eqs.\ (\ref{eq: self-consistency-full}) 
allow for a qualitative analysis of the
influence of spin-spin and dipole-dipole correlations 
on the elastic distortions.
We later provide the numerical evidence for the qualitative outcoming picture.

\begin{figure}
	\begin{centering}
		\includegraphics[scale=0.45]{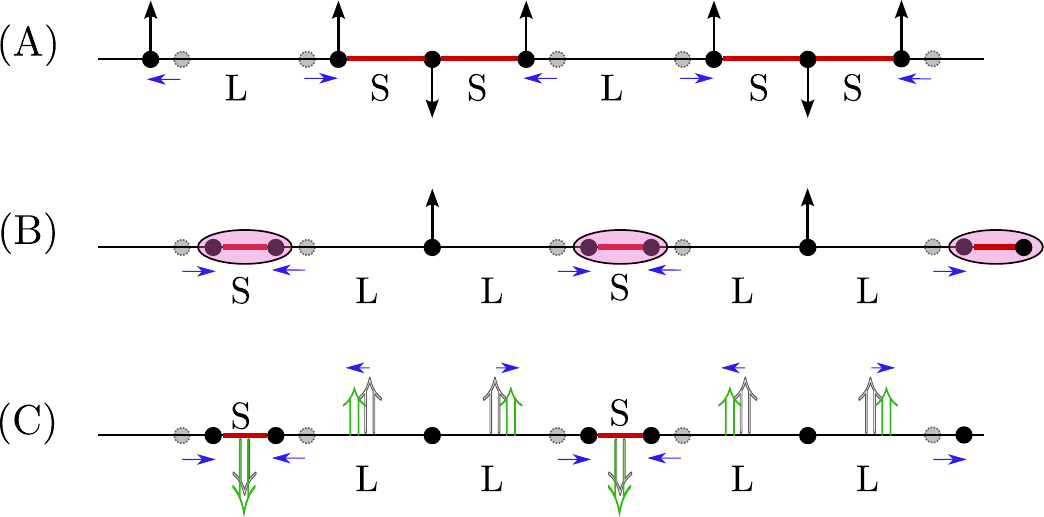}
		\par\end{centering}
	\caption{\label{fig: simple patterns}Qualitative pictures of:
		(A) a classical $\uparrow\uparrow\downarrow$ $M=1/3$ magneto-elastic state.
		(B) a quantum $\bullet\!\!-\!\!\bullet\uparrow$ $M=1/3$ magneto-elastic state.
		(C) the $P=1/3$ electro-elastic state in Fig.\ \ref{fig: extended electroelastic}.
		Black dots represent the lattice sites, black arrows the up/down spin states, ellipses the dimer spin singlet states, 
		and green double arrows the electric dipoles. 
		Non distorted sites (and dipoles in (C)) are indicated with gray faded symbols to appreciate ion displacements (blue arrows). 
		L and S indicate long/short bonds.
		 Enhanced NN ($J_1$) exchange couplings are highlighted in red.
		Notice that electro-elastic distortions in (C) are compatible with the quantum magneto-elastic ones in (B) 
		but not with the classical magneto-elastic ones in (A).	
	}
\end{figure}

Let us summarize more technically some pertinent results on magneto-elastic chains.
$M=1/3$ magnetic plateaus come in two flavors, dubbed classical and
quantum. \cite{Hida-Affleck-2005} 
In the so called classical
plateau spin components parallel to the magnetic field have non vanishing
$\langle S_{i}^{z}\rangle$ expectation value in an ordered pattern
with two positive, one negative terms that we represent by $\uparrow\uparrow\downarrow$.
These expectation values are reduced by quantum fluctuations in the
isotropic $\gamma = 1 $ case, but approach $\pm0.5$ in the highly easy-axis
anisotropic case $\gamma \ll 1 $. 
Spin-spin correlations $\langle\boldsymbol{S}_{i}\cdot\boldsymbol{S}_{i+1}\rangle$
are positive between ferromagnetic (parallel) neighbors $\uparrow\uparrow$
and negative between antiferromagnetic (antiparallel) neighbors $\uparrow\downarrow$
and $\downarrow\uparrow$, approaching the Ising correlations $\pm0.25$
for $\gamma \ll 1 $. 
From Eq.\ (\ref{eq: self-consistency-full}), the correlation
$\langle\boldsymbol{S}_{i}\cdot\boldsymbol{S}_{i+1}\rangle$ affects
the bond distortion $\delta_{i}$; 
the $\uparrow\uparrow\downarrow$
spin configuration favors distorted long bonds between ferromagnetic
neighbors and short bonds between antiferromagnetic neighbors, 
that
is a \textquotedbl long-short-short\textquotedbl{} (L-S-S) distortion
pattern (see Figures \ref{fig: pantograph} and \ref{fig: simple patterns}-A).
Notice that the antiferromagnetic coupling $J_{1}\left(1-\alpha\delta_{i}\right)$
gets stronger for \textquotedbl satisfied\textquotedbl{} antiferromagnetically
aligned neighbors and weaker for \textquotedbl frustrated\textquotedbl{}
ferromagnetically aligned neighbors. 

In contrast, in the so called quantum plateau two neighboring spins
(out of every three) tend to form singlets while the third one points up,
in a configuration that we represent by $\bullet\!\!-\!\!\bullet\uparrow$
(see Figure \ref{fig: simple patterns}-B). In an ideal case the spins
forming a quantum singlet would have $\langle S_{i}^{z}\rangle=0$
and the third one $\langle S_{i}^{z}\rangle=0.5$, with singlet correlation
$\langle\boldsymbol{S}_{i}\cdot\boldsymbol{S}_{i+1}\rangle=-0.75$
and vanishing correlation between the spin up and its neighbors; the
real situation may be characterized as a quantum plateau when the
spin expectation and spin-spin correlation values show a tendency
to such pattern. Again from Eq.\ (\ref{eq: self-consistency-full}) one
can see that a very negative singlet-like correlation strongly favors
a short bond at the expense of long bonds (according to Eq.\ (\ref{eq: basic constraint}))
where spin correlations are close to zero, giving rise to a \textquotedbl short-long-long\textquotedbl{}
(S-L-L) distortion pattern. Notice that the singlets are more likely
to appear in the isotropic case $\gamma=1$, while the easy-axis anisotropy
$\gamma<1$ diminishes transverse correlations and favors the classical
configuration. 
We must stress that it is the classical order the one
usually  observed in homogeneous $J_{1}-J_{2}$ magnetically frustrated spin
chains in a wide variety of regimes, either isotropic (with \cite{Vekua-etal-2006,Tertions}
or without \cite{2003-Okunishi-a} elastic coupling) or
anisotropic \cite{2003-Okunishi-b}. 

In turn, the NN dipolar correlations are related to lattice distortions
through the second line of Eq.\ (\ref{eq: self-consistency-full}): bond
distortion $\delta_{i}$ is influenced by the correlations of the
dipole $\sigma_{i}$ located at the bond $i$ with NN dipoles at both
sides. The $\Uparrow\Uparrow\Downarrow$ configuration then favors
short bonds where the dipole $\Downarrow$ is located, at the expense
of generating long bonds where the dipoles point $\Uparrow$ to fulfill
the constraint in Eq.\ (\ref{eq: basic constraint}), preferring to induce
a S-L-L distortion pattern (see Figure \ref{fig: simple patterns}-C).
Recalling that dipoles remain always midway between adjacent magnetic
atoms, in terms of dipole positions these magnetic lattice distortions
make antiparallel dipoles get closer, and parallel dipoles get further
away. The NNN dipolar interactions, necessary to introduce the $\Uparrow\Uparrow\Downarrow$
order in the electro-elastic phase diagram,
enter in Eq.\ (\ref{eq: self-consistency-full}) with smaller coefficients.

From this qualitative discussion, the electro-elastic dipolar configuration
$\Uparrow\Uparrow\Downarrow$ found in Fig.\ \ref{fig: extended electroelastic} is
compatible with the quantum magnetic plateau configuration but competes
with the classical plateau configuration.

%


\subsection{Numerical self-consistent analysis}

The coupling to dipolar degrees of freedom  introduces
a second frustration mechanism in the $J_1-J_2$ antiferromagnetic chain, 
favoring the stabilization of 
the elusive quantum order in the $M=1/3$ plateau state.
However, the classical plateau state is the lowest energy configuration 
generally found in most investigations \cite{Vekua-etal-2006}. 
An exception has been presented by tailoring inhomogeneous period three exchange couplings \cite{Hida-Affleck-2005}.
Wether the quantum or classical plateau shows up in the present case will depend on the several parameters, 
in particular 
the degree of frustration $J_2/J_1$, 
the incidence of quantum fluctuations governed by the anisotropy $\gamma$
and the strength of the magneto-elastic coupling $\alpha$ and electro-elastic coupling $\beta$.

The numerical self-consistent exploration of the extended pantograph model  
in the double frustration scenario \cite{pantograph-3}
shows that both the classical and quantum orders can be stabilized. 
Assuming magnetic and electric fields driving the system to magnetization
$M=1/3$ and polarization $P=1/3$
the role of magnetic frustration and easy-axis anisotropy  
has been explored to produce a diagram in  the $J_{2}/J_{1}-\gamma$
plane. The remaining parameters have been fixed 
with $\alpha=\beta$ to enable magneto-electric competition and
$J_{e}$ in the range where $P=1/3$ is observed
(see Figure \ref{fig: extended electroelastic}).

\begin{figure}
	\centering{}
	\includegraphics[scale=0.7]{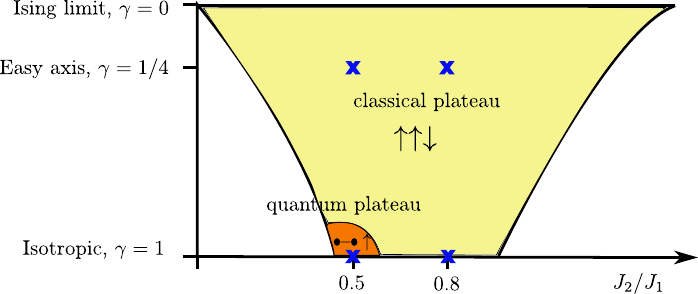}
	\caption{\label{fig: phase-diagram-DF}
		Schematic phase diagram in the frustration
		ratio ($J_{2}/J_{1}$) and the easy-axis anisotropy ($\gamma$) plane.
		The colored regions indicate the parameter regimes where $M=1/3$
		plateaus are observed in  magnetization curves (see Fig.\ \ref{fig: H vs J2/J1}). 
		The robust magnetic
		order giving rise to the plateau is mostly a collinear $\uparrow\uparrow\downarrow$
		classical structure (yellow region) but turns into a quantum $\bullet\!\!-\!\!\bullet\uparrow$
		state (orange region) for low frustration and small anisotropy. 
		Blue crosses mark the points explored in Ref.\ [\onlinecite{pantograph-3}].
	}
\end{figure}

The distinct regimes  discussed in Ref.\ [\onlinecite{pantograph-3}]
are:
\begin{enumerate}
	
	\item $\gamma=1$, $J_{2}/J_{1}=0.5$. Due to the isotropic Heisenberg interaction
	and the high magnetic frustration ($J_{2}/J_{1}=0.5$ is the maximally
	frustrated point in the case of Ising interactions) quantum fluctuations
	are enhanced at this point. 
	
	\item $\gamma=1/4$, $J_{2}/J_{1}=0.8$. Easy-axis anisotropy and low magnetic
	frustration inhibit quantum fluctuations, probably favoring classical behavior.
	
	\item $\gamma=1$, $J_{2}/J_{1}=0.8$,  a point with isotropic
	Heisenberg interaction and low frustration. 
	and low quantum fluctuations.
	
	\item $\gamma=1/4$, $J_{2}/J_{1}=0.5$, selected as a high magnetic frustration
	point with low quantum fluctuations.
	
\end{enumerate}
The formation of different $M=1/3$ plateau structures in different regimes  
is shown in Fig.\ \ref{fig: phase-diagram-DF}, 
where from the explored points we 
draw a schematic phase diagram.

\begin{figure}
	\centering{}
	\includegraphics[scale=0.5]{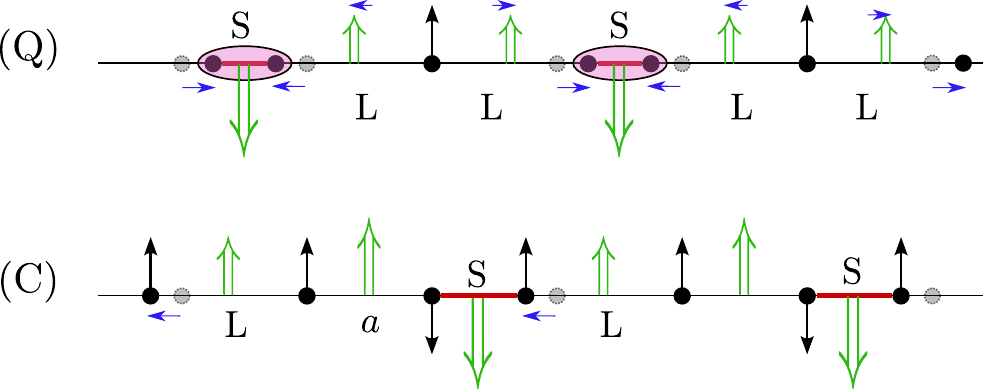}
	\caption{\label{fig: quantum-classical-M_1_3}
	Picture of the quantum (Q) and classical (C) $M=1/3$ configurations.
Symbols follow the conventions in Fig. \ref{fig: simple patterns}.
The displayed structures reflect actual data in Ref.\ [\onlinecite{pantograph-3}].
	}
\end{figure}

			
\subsection{Multiferroic effects at finite fields \label{sec: finite M effects}}

\subsubsection{Effects of a magnetic field on the polarization} 

According to the general philosophy of the pantograph model,
when a magnetic field drives the magneto-elastic subsystem into or out of 
a robust plateau configuration there is a sharp response from the electro-elastic sector.
We describe here a scenario for such a magnetic driven polarization jump 
at finite magnetization and polarization. Such kind of situations have been observed experimentally in compounds like LiCuVO, etc.
  
Consider the action of a fixed electric field setting the $\Uparrow \Uparrow \Downarrow$ 
configuration for the dipolar sector, 
and a magnetic field slightly below the range that supports the $M=1/3$ magnetization plateau.
The magnetic sector would be disordered (in a Luttinger liquid state \cite{Giamarchi}), 
so that the elastic distortions average to zero at each bond,   
the dipole strengths average to $p_0$ (see the early Eq.\ (\ref{eq: beta}))
and the polarization is $P=1/3$. 

\begin{figure}[ht!]
	\centering
	\includegraphics[width=0.75\columnwidth]{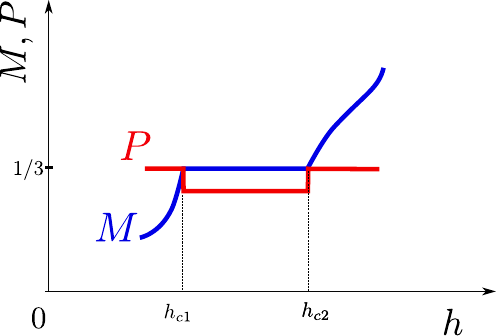} 
	\caption{Schematic description of the change in polarization driven by the magnetic field
		as it enters and leaves the double frustration plateau. 
	}
	\label{fig: Delta-P-un-tercio}
\end{figure}

When the magnetic field is raised, the magnetic sector adopts an ordered plateau configuration,
the distortions also get ordered and the dipoles have different strength in different bonds.
Now the net polarization is modified to $1/3+\Delta P$. According to Fig. \ref{fig: Delta-P-un-tercio}
one finds $\Delta P<0$ either when the parameters favor the quantum plateau configuration
or the classical one.

Once the magnetic field grows beyond the $M=1/3$ range the magnetic sector
gets progressively disordered because of the proliferation of solitons (domain walls).
While local polarization could be present, the bulk polarization will again average to $P=1/3$.

This scenario is sketched in Fig.\ \ref{fig: Delta-P-un-tercio}.
We recall that it is the polarization change the quantity that one can access experimentally. \cite{Resta-1993}

\subsubsection{Effects of an electric field on the magnetization} 

As described  in Sections \ref{subsec: M-vs-P} and \ref{subsec: extended M-vs-E}, 
a variation of the electric field driving the magneto-elastic sector 
out of a plateau  generally modifies the magnetization curve and 
habilitates a magnetization jump driven by the electric field.
We argue that this also occurs when the electric field  
takes the system out of the $P=1/3$ $\Uparrow \Uparrow \Downarrow$ configuration. 
Though the $M=1/3$ plateau is expected to remain because of magnetic frustration, 
it would be stabilized in a different magnetic field range. 
One could then fine-tune the parameters to find a region where,
at fixed magnetic field, 
a variation of the electric field would produces a finite jump 
$\Delta M$ in the magnetization.

			
\section{Higher dimensions and higher spins \label{sec: higher dimensions and spins}}

As mentioned in the Introduction, some further generalizations are needed in order to provide more precise models 
for the wide variety of existing multiferroic materials. In this Section we summarize our results for two important generalizations of the pantograph model: 
we first treat the one dimensional magneto-elastic model with higher spin magnetic ions, $S \geq 1$, 
and show that the pantograph mechanism could take place in a certain region of the phase space. 
Second, a two-dimensional version of the pantograph model is studied, in this case with Ising magnetic moments on a square lattice geometry.

\subsection{Spin $S>1/2$ magneto-elastic chains \label{sec: higher spins}} 

Most of the results described above rely on the existence of robust plateau states in 
spin $S=1/2$ chains. 
By robust we mean that gaps in the magnetization spectrum are significant with respect to 
the saturation Zeeman energy, but also that they do not need fine tuning of parameters. 
Indeed, as we have seen, the plateau states are observed in quite different Hamiltonians 
(homogeneous or modulated couplings, frustrated or not frustrated, isotropic or anisotropic interactions).
Extensions to higher spin naturally start by exploring the existence of spin gaps in the corresponding magneto-elastic spin chains.

\subsubsection{Model and methods}

We consider a simplest magneto-elastic spin chain with local spins $S \geq 1$, akin to the magneto-elastic sector in Section \ref{sec: minimal}. 
The magnetic interactions are described by 
nearest-neighbor antiferromagnetic exchange with Hamiltonian given by
\begin{eqnarray}
	H_\text{spin}^{(S)}=\sum_i J_i \mathbf{S}_{i}.\mathbf{S}_{i+1} - h \sum_i S^z_{i}. 
	\label{Hmag}
\end{eqnarray}
The minimal coupling to distortions modulates the exchange couplings as
\begin{eqnarray}
	J_i = J \left[ 1-\alpha (u_{i+1}-u_{i} )  \right]
	\label{J_j}
\end{eqnarray}
where we use $u_{i}$, the displacement of ion $i$ from its regular lattice position. 
That is, as before, $J_i$ depends linearly on the bond distortion $\delta_i = u_{i+1}-u_{i}$ 
through a spin-phonon coupling $\alpha$.
The lattice degrees of freedom are described by their energy cost in the adiabatic limit, 
under the assumption that phonon frequencies are much smaller than $J$.
However we introduce a difference here: for the elastic energy
we choose the so-called Einstein-site phonon (ESP) model \cite{Balents_2006} which considers
a quadratic energy arising from displacement of magnetic ions from their equilibrium positions 
(in absence of magnetic interactions),
\begin{eqnarray}
	H_\text{ESP}^{(S)}=\frac{K}{2}\sum_i u_{i}^{2}
	\label{J_i}
\end{eqnarray}
describing a dispersionless optical phonon branch. 
This choice provides similar results than Eq.\ (\ref{eq: H elastic}) but is more convenient for computational purpose in the present context.

This model could be extended, as before, by including frustration through NNN exchange $J_2$ or anisotropic interactions. 
Most interesting, for $S\geq 1$ it allows for single ion anisotropy \cite{Yosida-1968} terms with the usual form $D_i \left( S_i^z\right)^2$.
Such anisotropy, arising from spin-orbit coupling of the unpaired electrons in magnetic ions, has been considered to play 
an important role in the magnetic ordering of several type II multiferroic materials [REFERENCES].
Notice that such term is not relevant for $S=1/2$ ions.

The ground state of the magneto-elastic Hamiltonian $H_\text{spin}^{(S)} + H_\text{ESP}^{(S)}$ can be solved by the 
self-consistent method in Section \ref{sec: self consistency}, leading to no distortions or at most to alternate distortions 
(let us call $\delta_0$ the distortion amplitude)
Then the analysis of the magnetic sector can be done in the realm of modulated exchange antiferromagnetic spin chains 
with alternate couplings $J_i=J(1-(-1)^i 2 \delta_0)$ (the so-called dimerized spin chain).

The purely magnetic sector with spin $S$ can be treated analytically  by the bosonization of $2S+1$ locally coupled spins $1/2$ 
(leading to $2S+1$ coupled Abelian bosonic fields) \cite{Schulz}.
Confirming the well-known Haldane conjecture, \cite{Haldane_1983} 
the outcome is qualitatively very different according to $S$ being integer or half-odd-integer.
On the one hand the integer spin chains always exhibit a zero magnetization gap, known as the Haldane gap;
they become magnetized only when an external magnetic field exceeds a threshold value.
On the other hand the half-odd-integer spin chains are gapless.

Regarding the existence of a spin-Peierls gap, leading to spontaneous dimerization, it is theoretically  expected for half-odd-integer spins.
Based on the low-energy bosonic field theory mapping \cite{Schulz} it was conjectured \cite{Affleck-Haldane-1987} 
that the spin-Peierls instability should also take place in chains with half odd-integer spin greater than $S\geq 3/2$. 
This has been partially supported from numerical efforts \cite{Affleck_1989} 
and theoretically using the non-Abelian bosonization mapping \cite{NAbos}.
But in spite of the theoretical efforts, numerical evidence is hardly conclusive for chains with spin higher than 1/2.

Motivated by the widespread significance of the spin-Peierls transition in magneto-elastic systems, 
by the expected differences between integer and half-integer spin chains, 
and the by lack of conclusive prior studies on systems with $S > 1$, 
we have revisited the $S=1$ and the $S=3/2$ cases applying 
the numerical (DMRG) self-consistent method described in Section \ref{sec: self consistency}. 



\subsubsection{Results}

Analyzing the $S=3/2$ model in absence of magnetic field for a wide range of the dimensionless spin-phonon coupling $\lambda\equiv J\alpha^2/K$,
aside from the elusive question of the existence of a spin-Peierls instability for weak $\lambda$, 
we find a feature not present in the $S=1/2$ physics which is the main message in this Section: 
a first-order structural transition as 
a function of the spin-phonon coupling from a homogeneous/weakly-dimerized  phase with antiferromagnetic exchanges at low coupling
(which would be consistent with the field theoretical expectations \cite{Schulz, Affleck-Haldane-1987, NAbos})
into a strongly dimerized phase with alternating ferro- and antiferro-magnetic exchange interactions (dubbed here as FM-AFM phase) realized at strong coupling. 
The two distinct regimes could be observed in different materials, according to their intrinsic spin-phonon coupling, but more interestingly
the transition could be driven  by a variety of experimentally controllable parameters such as striction, magnetic or electric fields, etc. 



\begin{figure}[ht!]
	\centering
	\includegraphics[width=1.0\columnwidth]{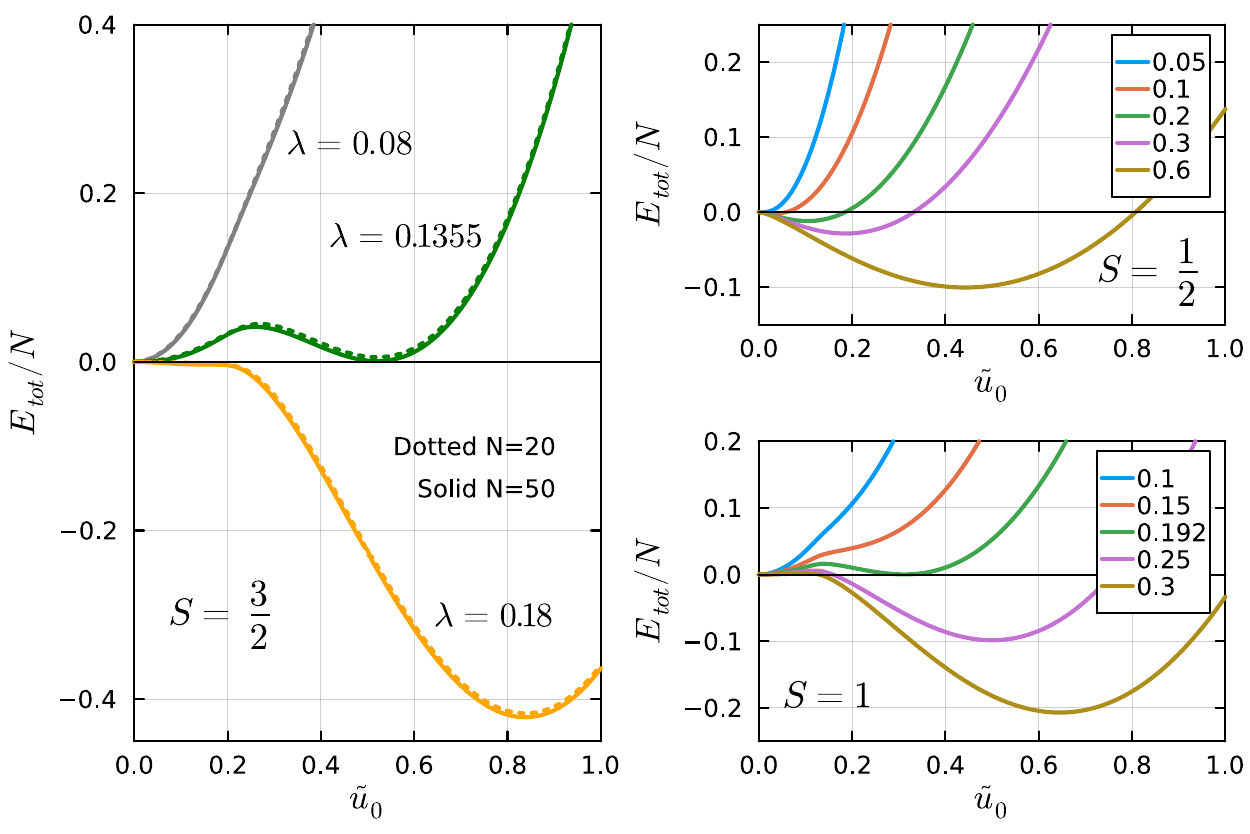} 
	\caption{Left: total energy computed for alternating displacements as a function of their amplitude $\tilde{u}_0$
		for $S=3/2$ and different values of the spin-phonon coupling $\lambda$. Right top (bottom): the same for $S=1/2$ ($S=1$) and different $\lambda$ indicated in the insets. (Reprinted from [\onlinecite{Spin-Peierls-2024}])  }
	\label{fig: P4-energy-landscape}
\end{figure}

Starting with $h^z=0$, for the full range of the spin-phonon coupling $\lambda$ we find
that the ground state is a spin singlet ($S^z_{tot}=0$) and exhibits a period 2 pattern of ion displacements, 
say $\tilde{u}_i=(-1)^i \tilde{u}_0$, producing an alternation of short and long bond distances.  
Depending on $\lambda$ we have found two strikingly different solutions, 
as can be seen in Fig.\ \ref{fig: P4-energy-landscape}, where we show 
the total (magnetic plus elastic) energy for dimerized distortions 
as a function of the distortion amplitude 
$\tilde{u}_0$
for different values of $\lambda$.

In the weak coupling regime (lowest $\lambda$ in the figure) it is  difficult to distinguish 
whether the minimum energy is obtained for a homogeneous configuration or for a slight dimerization. 
Numerically we assume that the chain shows no distortions. 
In contrast, for strong coupling (largest $\lambda$ in Fig.\ \ref{fig: P4-energy-landscape})   
a highly distorted phase shows up with $\tilde{u}_0$ of the order of 0.5 
In the latter case physically meaningful distortions ${u}_i = \tilde{u}_i/ \alpha$, 
which should be much smaller than  the lattice spacing $a$,  require materials with $\alpha \gg 1/a$.
At some critical value of $\lambda$ a 
first-order transition takes place.
The transition to the strongly dimerized behavior 
corresponds to a second local energy minimum becoming more favorable than the expected homogeneous or weakly-dimerized configuration,
as illustrated in Fig.\ \ref{fig: P4-energy-landscape} by results at, below and above the critical value, 
estimated as $\lambda_c \approx 0.1355$. 
Energies do not scale significantly with the chain length.
One can see the presence of two local minima, one for vanishing or tiny $\tilde{u}_0$ and another for large $\tilde{u}_0$.
The second one becomes energetically favorable at strong coupling 
$\lambda$.
We then identify a level crossing, a quantum first-order transition as a function of the spin-phonon parameter 
$\lambda$. 

To stress the finding, we show at the right of Fig.\ \ref{fig: P4-energy-landscape} the corresponding results for lower spin.
Clearly, for $S=1/2$ the first-order transition is absent (right top panel).
For $S=1$ a first-order transition could be expected due to the existence of a finite Haldane gap in integer $S$ spin chains. 
Our data (right bottom panel) agree with previous ones. 
In particular, the critical coupling we find for the transition $\lambda_c\approx 0.192$ is the same as that reported in Ref. [\onlinecite{Onishi02}]. 
Note that this value is near 42\% larger than the one we find for $S=3/2$. 
Therefore, the transition should be easier to observe in $S=3/2$ chains. 
In addition, the effects of an applied magnetic field should be easier to measure for larger spin magnetic ions.

It is noticeable that contrary to naive expectations, the qualitative behavior for the $S=3/2$ chain more closely resembles
that of the  $S=1$ case, rather than the other half-integer spin system $S=1 /2$.


\begin{figure}[ht!]
	\centering
	\includegraphics[width=1.0\columnwidth]{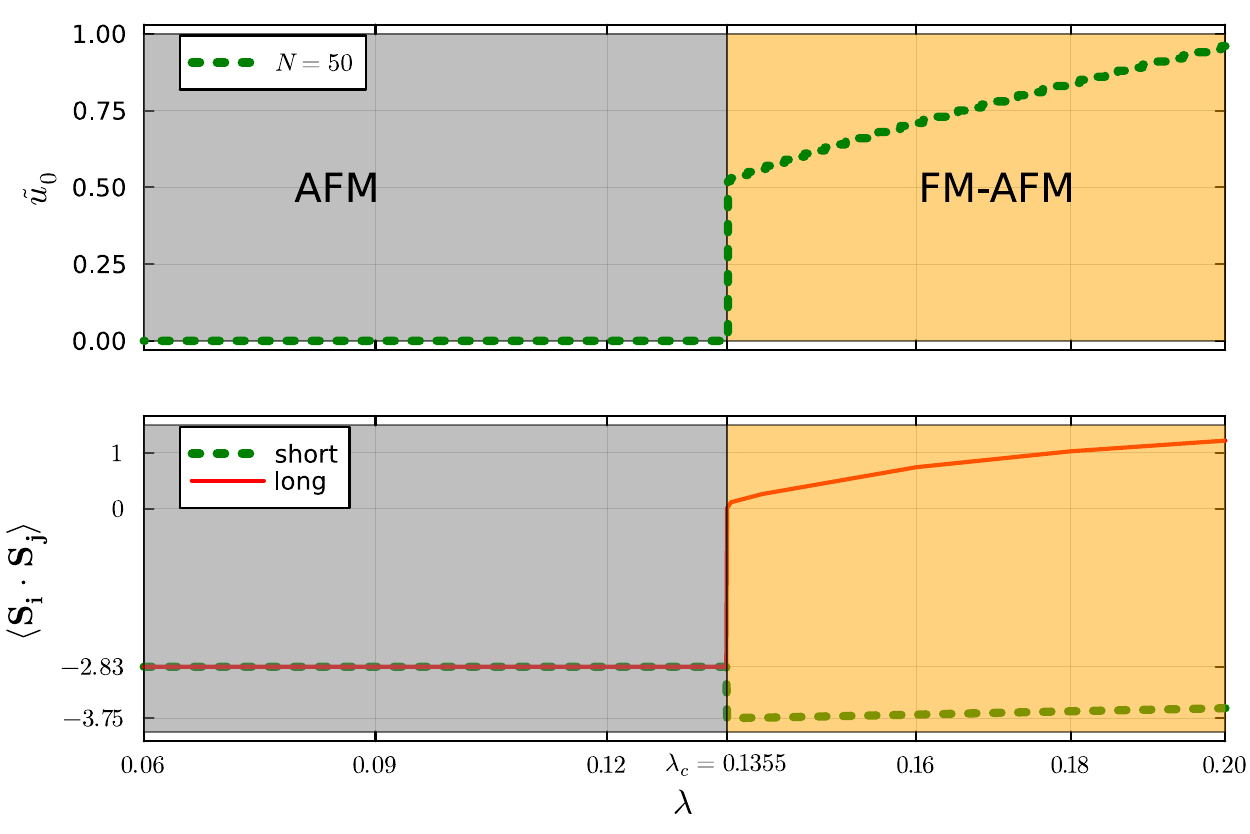} 
	\caption{Amplitude of alternating displacements (top) and spin-spin correlations (bottom) in the ground-state  as a function of the spin-phonon coupling $\lambda$.  (Reprinted from [\onlinecite{Spin-Peierls-2024}])}
	\label{fig: P4-distortion-amplitude}
\end{figure}

In Fig.\ \ref{fig: P4-distortion-amplitude}, top panel, we plot the amplitude of alternating displacements versus $\lambda$
in the ground state of the system, for a sample chain length. 
One can clearly see a jump, occurring at 
$\lambda_c \approx 0.1355$. 
Within our numerical precision the value of 
$\lambda_c$ is not sensitive to the chain length, 
a robust feature that should be valid in the thermodynamic limit.  
Beyond the critical point the displacement amplitudes jump to values larger than 0.5, 
making the exchange $J_i = J \left[ 1-(\tilde{u}_{i+1}-\tilde{u}_{i} )  \right]=J ( 1-2\tilde{u}_0)$  
small and negative at bonds where ions become separated. 
This generates an important alternation of the spin exchange between 
strong antiferromagnetic dimers (at short bonds) and weaker ferromagnetic interactions (at long bonds), 
producing a deep effect in the correlations in the strong coupling phase. 
The nearest neighbors spin-spin correlations alternate between  short and long bonds:
at each short bond the correlation indeed takes a value close to 
$-\frac{3}{2}\left(\frac{3}{2}+1\right)=-\frac{15}{4}$, which is the value corresponding to perfect singlet dimer states. 
For the long bonds the spin-spin correlations are in general positive (ferromagnetic), as shown in 
Fig.\ \ref{fig: P4-distortion-amplitude}, bottom panel.
Importantly, at the transition we find $\tilde{u}_0=0.5$, 
so that perfect singlet dimers form in the AFM bonds with correlations  $-\frac{15}{4}$, 
exactly decoupled from each other.

The dimerization of exchange couplings produces a finite spin gap,
defined as the difference of the total (magnetic plus elastic) energy 
between the lowest lying states with $S^z_\text{total}=1$ and $S^z_\text{total}=0$.

Given the abrupt onset of alternating distortions described in the previous Section, 
the spin gap indeed jumps from zero in the AFM phase
to a finite value in the FM-AFM phase. 

\begin{figure}[ht!]
	\centering
	\includegraphics[width=1.0\columnwidth]{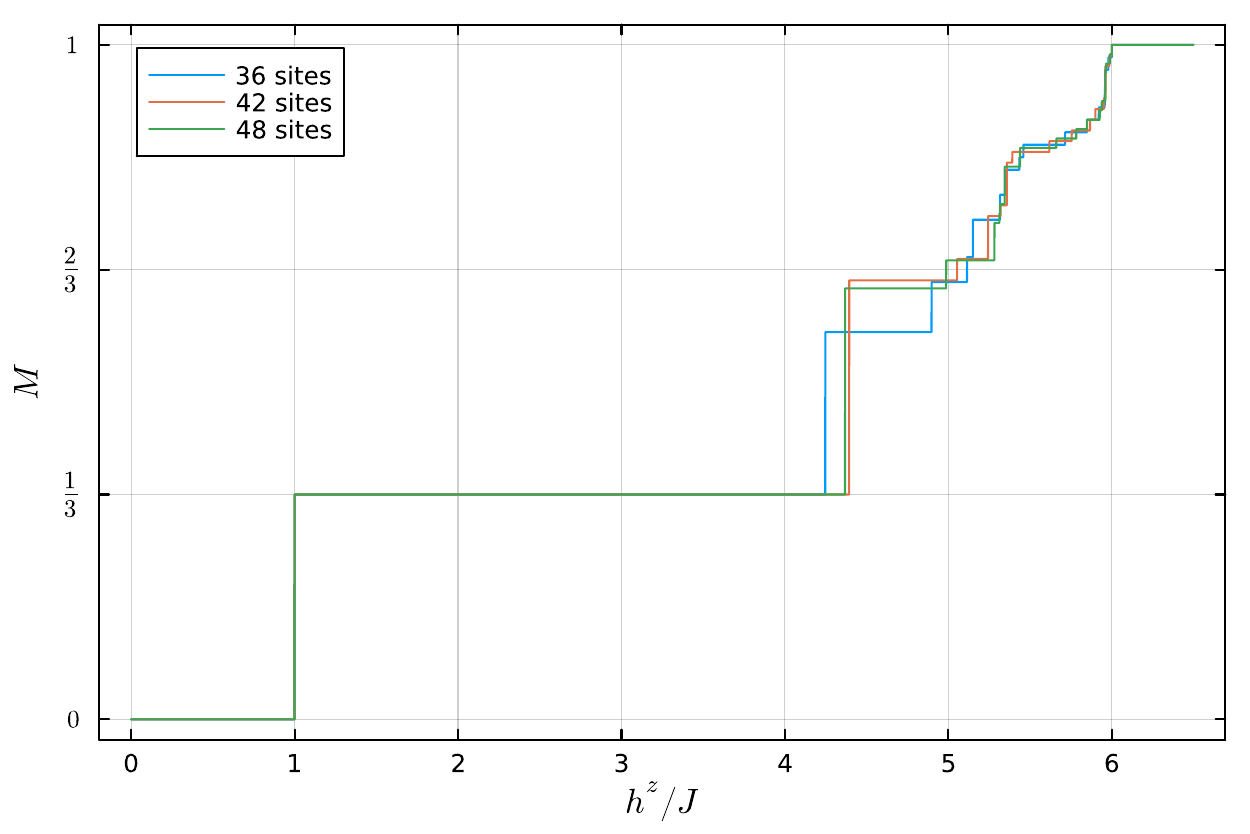} 
	\caption{Magnetization curves of the magneto-elastic model, for $\lambda=0.18>\lambda_c$ 
		and three different chain lengths  (Reprinted from [\onlinecite{Spin-Peierls-2024}]). 
		}
	\label{fig: P4-magnetization-curve}
\end{figure}

The magnetization curve is smooth for low spin-phonon coupling (AFM phase) but presents robust plateaus
when a large enough spin-phonon coupling sets the system in the FM-AFM phase. 
A test case is shown in Fig.\ \ref{fig: P4-magnetization-curve}, with plateaus at
$M=0$ and $M=1/3$.

\begin{figure}[ht!]
	\centering
	\includegraphics[width=0.9\columnwidth]{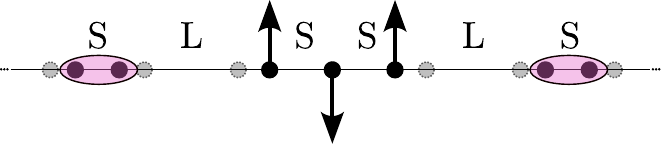} 
	\caption{Elementary magnetic excitation of the $S=3/2$ magneto-elastic chain. 
	Arrows here represent $S^z _i=3/2$  spin states and ellipses represent $S=3/2$ dimer singlets.
	Letters  S, L emphasize the alternation  of short, long  bonds disrupted by a domain wall.
	}
	\label{fig: 3spinon}
\end{figure}

Exiting from the $M=0$ plateau (with alternate distortions) the lowest magnetic excitation has not $S^z_\text{total}=1$ 
but instead $S^z_\text{total}=3$, 
that would correspond to a rigid flip of a $S=3/2$ classical vector.
This excitation decouples into two  localized solitons (domain walls) depicted in Fig.\ \ref{fig: 3spinon}:
each one appears to be localized in a range of three sites, close together, 
with spin projections $+\frac{3}{2},-\frac{3}{2}, +\frac{3}{2}$, carrying spin $3/2$. 
Such solitons separate regions in a sea of singlet dimers, 
differing by spatial translation of one lattice site. 

From the distortion patterns decribed for the $S=3/2$ magneto-elastic chain
it is clear that a pantograph coupling to dipolar degrees of freedom generates a net ferrielectric polarization
at zero magnetic field, and that the polarization will switch-off when the magnetic field exceeds a threshold.
The remarkable property is that the mechanism itself could be set active/inactive 
by controlling the spin-phonon coupling $\alpha$  ($\lambda$ in dimensionless language).  
This control could be implemented by some external knob, most likely elastic striction by application of pressure. 

The uncovered FM-AFM phase transition is potentially relevant for multiferroic applications whenever a tailoring of $\lambda$ leads to a giant variation of the magnetic susceptibility. 
It promises prospective generalizations of the pantograph model and associated ferroelectric effects.
The presence of the first-order transition, and the associated large change in the crystal structure, 
could produce a noticeably drastic jump in both magnetization and electric polarization 
in improper type II multiferroic materials if spins $S>1$ are involved. 
Although a similar first-order transition has been observed for $S=1$, 
it has a different origin and the critical value of $\lambda$ required to induce it is much higher than that for $S=3/2$.


\subsection{A two dimensional pantograph model \label{sec: Pili}}

The basic idea of the pantograph mechanism, 
that is the association of active electric dipolar moments to lattice distortions, 
has been applied by one of the authors and collaborators to a magneto-elastic Ising model in the square lattice. \cite{Pili-Cabra}
As compared with previous works where the dipole-dipole interaction was not taken into account, \cite{Pili-2020} 
we should stress that such coupling stabilizes the 
$\uparrow \uparrow \downarrow \downarrow$ 
magnetic ordering along zig-zag stripes. 
The model successfully reproduces the same key phenomena as its one-dimensional counterpart, 
that is the sudden drop-off of the bulk electric polarization as a function of the magnetic field, simultaneously with a sudden increase of the magnetization, together with the 
$\uparrow \uparrow \downarrow \downarrow$ 
ordering observed in several type-II multiferroic materials (see {\it e.g.} \cite{Dagotto-2006,Dong-2009,Tokura-2009}).


\subsubsection{Model and methods}

The magneto-elastic sector is described by the Hamiltonian
\begin{equation}
	H_{ME}^\text{Ising} = \sum_{\langle i,j\rangle} J(r_{ij}) S_i S_j+\frac{K}{2}\sum_i \delta_{ij}^2,
	\label{H magneto elastic Pili}
\end{equation}
where $S_i=\pm 1$ stands for Ising type variables (magnetic moments pointing along a preferred lattice direction) 
at sites $i$ in a square lattice and $ \langle i,j\rangle$ indicates nearest neighbors sites.  
More precisely we call  $\mathbf{r}_i^0$ the position of magnetic ions in the regular lattice and 
$\delta \mathbf{r}_i $  their displacements. The exchange couplings depend on distortions as
\begin{equation}
J(r_{ij}) = J_0 (1-\alpha \delta r_{ij})
\label{H Jij Pili}
\end{equation} 
with $\delta r_{ij}=| \delta \mathbf{r}_j - \delta \mathbf{r}_i |$
measuring the change of distance between sites under distortions.

In contrast with previous sections, in the present scenario 
dipolar moments are not present when the lattice is not distorted. 
They arise at each site just because of distortions, and point along the local distortion directions.
They are represented by
\begin{equation}
	\mathbf{p}_i = \eta_i   \delta \mathbf{r}_i,
	\label{eq: p Pili}
\end{equation}
with a proportionality coefficient $\eta_i$ that could depend on the site type 
when the unit cell contains non-equivalent magnetic sites 
(eg. a bipartite square lattice).
This dipoles are active degrees of freedom, in the sense that they are coupled by the electrostatic dipolar interactions 
in Eq.\ (\ref{eq: dipolar interaction full}).

The ground state configurations, for different values of the parameters, were found by extensive simultaneous Monte Carlo simulations
of magnetic and elastic degrees of freedom, with dipoles following from elastic distortions.
Planar ion displacements were simulated in modulus below a reasonable cut-off and in angle by a detailed clock model. 
As displacements turn out to be always diagonal, a four state clock model was used for refined computations
(see [\onlinecite{Pili-Cabra}] for more technical details).

The action of external magnetic and electric fields have been considered \cite{Pili-Cabra}  adding to the Hamiltonian 
\begin{equation}
	H_\text{fields}=-h\sum_i S_i - \sum_i 
	\mathbf{E}\cdot \mathbf{p}_i
\end{equation}
where the magnetic field points along the Ising axis 
and the electric field was explored along both diagonal directions of the lattice.

 
\subsubsection{Results}

Antiferromagnetic ($J_0 >0$) and ferromagnetic  ($J_0 < 0$) couplings can be considered at once by flipping $S_i$ every two sites
(either even or odd sublattices). 
For convenience of description we refer below to the ferromagnetic case.

In absence of external fields three well defined distortion patterns  were found, 
at lowest temperatures achieved, for different model parameters. 

\begin{itemize}
	\item A regular lattice (no distortions) with ferromagnetic order.
	
	\item A checkerboard lattice where distortions lead to contracted square plaquettes. 
	The shortened  bonds in these plaquettes reinforce the ferromagnetic correlations, 
	while the couplings along the remaining elongated bonds change sign ($\alpha \delta r_{ij}>1$) 
	generating antiferromagnetic correlations. 
	
	\item Alternate distortions along one (spontaneous)  diagonal direction, 
	for instance an even sublattice displaced north-east and the odd  sublattice displaced south-west.
	This shortens the zig-zag bonds along all diagonals running north-west to south-east, reinforcing their ferromagnetic correlations.
	At the same time, distances between neighbors in contiguous zig-zag ferromagnetic stripes are augmented with ($\alpha \delta r_{ij}>1$),
	generating antiferromagnetic correlations. This configuration, dubbed zig-zag stripe, is depicted in Fig.\ \ref{fig: pili-patterns}.
	
\end{itemize}

Notice that the checkerboard and the zig-zag stripe configurations can show up only for large enough magneto-elastic
coupling $\alpha$. 
Indeed, they require   $\alpha \delta r_{ij}>1$ with reasonably small distortions $\delta r_{ij} \ll a$, 
not to break the lattice and  to support the linear expansions assumed in the computations.
They are energetically favorable against the uniform lattice when the spin-phonon coupling $\alpha$ exceeds some critical value, 
$\alpha > \alpha_c $. 
Once this happens, it is important to notice that both the checkerboard and zig-zag stripe configurations exhibit 
$\uparrow \uparrow \downarrow \downarrow$ 
magnetic order, along the horizontal and vertical lines of the square lattice. 
Moreover, the energy of these configurations is very similar.
If dipolar interactions are not included ($\lambda_D=0$), it was proven \cite{Pili-2020} that  
the checkerboard configuration is energetically favorable. 
But it was later shown  \cite{Pili-Cabra}  that when $\lambda_D$ is larger than some critical value $\lambda_D^c$ 
it is the stripe configuration the one with lowest energy. 
One should stress that the magnetic order in the zig-zag stripe is precisely the  E-type antiferromagnetic order observed 
in numerous multiferroic materials like 
TbMnO$_3$ and other RMnO$_3$ (R standing for rare earth elements) manganite perovskites and YNiO$_3$ nickelate perovskite.

The reason for the stability of the zig-zag stripe state
can be understood with the help of Fig.\  \ref{fig: pili-patterns}: because of zig-zag distortions
the dipoles pointing along the polarization axis (south-west to north-east)  are organized head to tail,
while dipoles along the transverse direction are antiparallel, gaining in both directions Coulomb dipolar energy. 
The dipolar interaction is the key ingredient for this explanation.

\begin{center}
	\emph{}
	\begin{figure}
		\begin{centering}
			\includegraphics[scale=0.8]{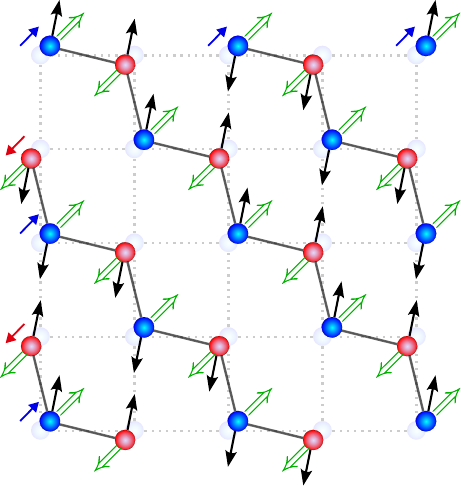}
			\par\end{centering}
		\caption{\label{fig: pili-patterns} 
		 Zig-zag stripe configuration, found to be the ground state in the two dimensional 
		ferromagnetic pantograph model when the spin phonon coupling $\alpha$  and 
		the dipolar interaction coupling $\lambda_D$ are larger than respective critical values. 
		Blue/red arrows shown on left and top borders indicate the displacements of blue/red sublattices of magnetic, charged ions
		with respect to the regular square lattice positions (in light gray).
		Ising spins, shown with black arrows, form ferromagnetic stripes along bonds where ion distances get shorter. 
		Parallel stripes, displaced away from each other, order antiferromagnetically.
		This crystallizes the observed $\uparrow \uparrow \downarrow \downarrow$ order along 
		the two directions of the square lattice.
		Dipolar moments, shown with green doble arrows, 
		are proportional to ion displacements adopting an antiferroelectic configuration.
		}	
	\end{figure}
	\par\end{center}

In a homogeneous square lattice with uniform polarization coefficient $\eta_i$ the zig-zag stripe distortions
produce  an antiferroelectric order without net polarization. 
Instead, in a bipartite lattice with different $\eta_i$ in even/odd sublattices, the same distortions
would lead to a ferrielectric state with net bulk polarization.
Perhaps more realistic, a charge disproportionation between even/odd sublattices would also produce a net polarization;
thus the model reproduces, from microscopic interactions, the phenomenological description 
of the multiferroicity observed in the perovskite  YNiO$_3$,  \cite{CheongMostovoy} among other materials.

In this case the model describes an improper type II collinear multiferroic transition. 
As the temperature is lowered from a disordered paramagnetic phase, one finds a simultaneous magnetic 
and structural transitions  with the byproduct of a ferrielectric configuration with net bulk polarization. 

Now, the action of an intense enough external magnetic field 
will destroy the antiferromagnetic zig-zag stripe order, leading the magnetic sector to full magnetization. 
Then the distortions disappear and the electric polarization is lost.
In this sense the two dimensional magneto-electro elastic Ising model predicts an electrical polarization switch-off controlled by a magnetic field.

This simple yet comprehensive model offers a framework for understanding the mechanisms 
behind magneto-electric coupling and phase transitions in type-II multiferroics, 
serving as a valuable tool for exploring low-energy device applications.


\section{Summary and perspectives \label{sect:conclusions}}

In a series of works we have developed a microscopic mechanism of magneto-electric coupling mediated by lattice distortions,
aimed to  building a realistic model for type II collinear multiferroic materials. 
At the root of the mechanism is the formation of magnetization plateaus in several magneto-elastic spin systems. 
Essential ingredients to match with experimental observations are the easy axis anisotropy $\gamma < 1$ 
favoring collinearity, the magnetic frustration $J_2/J_1$ leading to the
$\uparrow \uparrow \downarrow \downarrow$ 
spin ground state  
and the Coulomb-like long range dipole-dipole interaction establishing the antiferroelectric order, 
all of these in the absence of external fields. 

Motivated by the variety of known multiferroic materials,
which includes the $SU(2)$ symmetric as well as strongly easy axis anisotropic  spin interactions, 
we have explored the proposed model in several cases.
We cover, in one dimensional lattices,  from the spin isotropic regime $\gamma=1$  
up to  Ising-like anisotropic cases $\gamma \ll 1$, with or without magnetic frustration and different spin values.
We have also studied a two dimensional model, in this case for Ising magnetic moments. 

The microscopic mechanism may be described by a spin-dipole-Peierls Hamiltonian, 
where the indirect magneto-electric coupling  arises from a combination of a spin-Peierls like magneto-electric coupling,
which is known to lead to an elastic dimerization instability, 
and a pantograph mechanism that relates the strength of electric dipolar moments to lattice deformations.
Both mechanisms are ubiquitous in multiferroic materials, 
specially when competing magnetic interactions frustrate an antiferromagnetic N\'eel configuration.
Magnetic and electric degrees of freedom can thus either cooperate or compete in provoking lattice instabilities,
in a precise way expressed in the key self consistent  Eqs.\ (\ref{eq: self-consistency-full}).

We have shown, using complementary theoretical and numerical techniques,
that in a wide parameter region, starting at the isotropic $SU(2)$ Heisenberg model 
and going up to an extreme anisotropic ANNNI model, 
the system has a gapped magnetic ground state associated to dimerized lattice distortions.
Main consequences are the zero magnetization plateaus in the magnetization curves and the emergence of
an spontaneous ferrielectric bulk polarization (an antiferroelectric with a remanent polarization), 
with two possible degenerate orientations ($\mathbb{Z}_2$ symmetry).   

In the presence of an external magnetic field exceeding a critical value,
related to the spin gap,  low 
magnetization excitations develop as pairs of topological solitons 
that separate  different dimerized domains carrying opposite ferrielectric polarizations. 
A lattice of equidistant solitons grows along the system, 
producing a sharp switch off in the bulk polarization.
This mechanism, robust due to its topological character, 
could be at the root of the  bulk polarization jumps observed in many different
multiferroic materials. 
We expect that the present paradigm might be fitted to actual experimental parameters and 
be identified as one of the microscopic mechanisms behind magnetically induced polarization jumps.

We have also found a  polarization state at intermediate electric fields 
with $\Uparrow\Uparrow\Downarrow$ periodicity,
exclusively due to the long range character of the dipolar interactions 
frustrating the antiferroelectric order. 
Such a period three dipolar configuration, 
combined with the $M=1/3$ magnetic plateau state found at intermediate magnetic fields,
could give rise to interesting magneto-electric cross effects. This will be studied elsewhere.

Regarding technological interest, a material described by our model  
has a spontaneous $\mathbb{Z}_2$ polarization due to dipolar imbalance that can be easily controlled by applied fields.
In fact the presence of a small poling electric field 
gives rise to a relative displacement of the solitonic domain walls,
making the polarization of the magnetized states not to be completely turned off. 
Then a demagnetization would select a preferred orientation for the spontaneous polarization.
This property could be used, for instance, to engineer polarized memory storage devices controllable by very low electric signals. 
From a different point of view, the present work could guide the design and manufacture of composite artificial multiferroic systems, 
such as  multilayers (see for instance [\onlinecite{Taniyama-2015}]) 
where the mechanical strain transfer couples ferroelectricity and ferromagnetism, 
or even regularly nano-patterned arrays (see for instance [\onlinecite{Cai-2018}]) 
where flexoelectricity couples magnetostrictive strain gradients with electric polarization, in different materials.
The technological control of multiferroicity in these multiphase composite systems is rapidly progressing 
and could in a future be the alternative to chemically synthesized multiferroic compounds.
We hope that the understanding of the mechanisms of multiferroicity at the atomic scale 
will shed light on the effective magneto-electric coupling mechanisms taking place at the nanometer scale.

The pantograph mechanism, which is the key ingredient in our proposal to generate the magneto-electric coupling,  
encodes the relation between the dipolar moments and their lattice environment and 
is present as well in two or three dimensional systems. 
Appropriate extensions of the present model can be written taking into account detailed crystallographic data.

\section*{Acknowledgements}

The authors thank  A.A. Aligia, R.A. Borzi, A.O. Dobry, C.J. Gazza, S.A. Grigera, R.A. and L. Pili  
for their crucial collaboration in the publications that led to the present review.
This work was partially supported by CONICET (PIP 1146) and UNLP (PID X926), Argentina.


\end{document}